\documentclass[aps,prb,reprint,superscriptaddress,floatfix]{revtex4-1}
\usepackage[ngerman,english]{babel}

\usepackage{braket}
\usepackage{amsmath}
\usepackage[arrow, matrix, curve]{xy}

\usepackage{graphicx}
\usepackage[caption=false]{subfig}
\usepackage{wrapfig}
\usepackage{floatrow}

\makeatletter
\newcommand*{\rom}[1]{\expandafter\@slowromancap\romannumeral #1@}
\makeatother

\begin{document}


\title{Interaction-induced charge transfer in a mesoscopic electron spectrometer}




\author{Stefan G.~Fischer}
\affiliation{Department of Physics, Ben-Gurion University of the Negev, Beer-Sheva, 84105 Israel}
\affiliation{Department of Condensed Matter Physics, Weizmann Institute of Science, Rehovot, 76100 Israel}

\author{Jinhong Park}
\affiliation{Department of Condensed Matter Physics, Weizmann Institute of Science, Rehovot, 76100 Israel}

\author{Yigal Meir}
\affiliation{Department of Physics, Ben-Gurion University of the Negev, Beer-Sheva, 84105 Israel}

\author{Yuval Gefen}
\affiliation{Department of Condensed Matter Physics, Weizmann Institute of Science, Rehovot, 76100 Israel}


\date{\today}

\begin{abstract}
\noindent 
A novel mesoscopic electron spectrometer
allows for the probing of relaxation processes in quantum Hall edge channels. The device is composed of an emitter quantum dot that injects energy-resolved electrons into the channel closest to the sample edge, to be subsequently probed downstream by a detector quantum dot of the same type. In addition to inelastic processes in the sample that stem from interactions inside the region between the quantum dot energy filters (inner region), anomalous signals are measured when the detector energy exceeds the emitter energy.
Considering finite range Coulomb interactions in the sample, 
we find that energy exchange between electrons in the current inducing source channel and the 
inner region, similar to Auger recombination processes, is responsible for such anomalous currents. In addition, our perturbative treatment of interactions shows
that electrons emitted from the source, which dissipate energy to the inner region before entering the detector, 
contribute to the current most strongly when emitter-detector energies are comparable. Charge transfer in which the emitted electron is exchanged for a charge carrier from the Fermi sea, on the other hand, preferentially occurs close to the Fermi level.  
\end{abstract}

\pacs{}

\maketitle

\section{Introduction}

Coherent transport in chiral quantum Hall edge channels of mesoscopic devices is of considerable conceptual importance. Chiral channels enable the electronic implementation of originally optical interferometers, such as of the Fabry-Perot\cite{Ofek5276} or of the Mach-Zehnder\cite{ji1,PhysRevB.76.161309} type. In the electronic versions of such devices, interference occurs between the paths of the respective quasi-particles, which may enable the observation of anyonic statistics in the fractional quantum Hall regime.\cite{PhysRevB.55.2331,Han2016} Moreover, the implementation of quantum computational operations using edge channels is conceivable.\cite{RevModPhys.80.1083}

Relaxation dynamics of the electronic system plays a crucial role for transport properties of quantum Hall edges. 
Generically, relaxation processes exert a detrimental influence on coherence properties of edge channels, such as, e.g.,~on the interferometers' fringe visibility.~\cite{ji1,PhysRevB.76.161309,neder1,Freulon2015,Bocquillon2019}  
In addition to the fundamental interest in the phenomenon, it is therefore desirable to acquire a comprehensive picture of possible mechanisms which contribute to relaxation of non-equilibrium charge carrier distributions.

\begin{figure*}
  \subfloat[sample micrograph\label{fig:sample}]{%
    \raisebox{1.3cm}{\includegraphics[width=.44\textwidth]{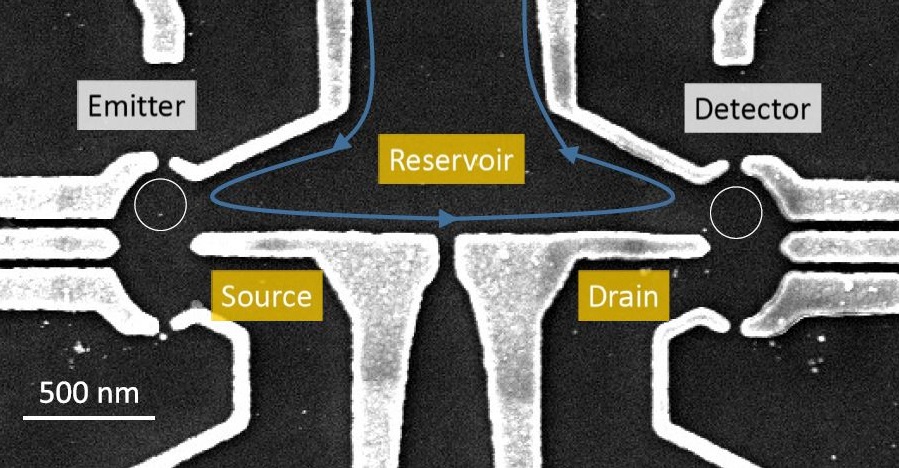}}
  }
  \hfill
      \subfloat[detector current\label{fig:current}]{%
    \includegraphics[width=.48\textwidth]{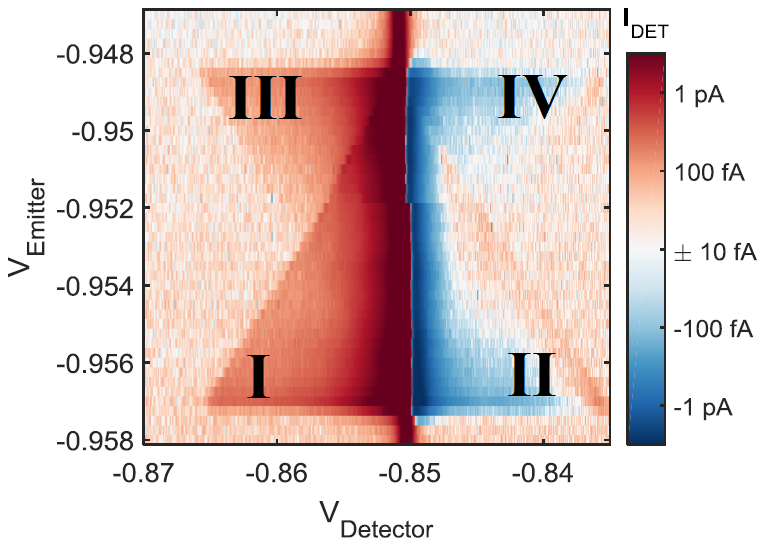}
  }
      \caption{(From Refs.~[\onlinecite{kraehenmann1}] and~[\onlinecite{arXivExp}]) (a) Mesoscopic sample for the spectral selection of charge carriers. Electrons from a source electrode are injected at a well-defined energy into a reservoir region via an emitter quantum dot. Subsequent energy selection at a detector quantum dot allows for the probing of intermediate relaxation of the electrons in the reservoir region. In the presence of a strong magnetic field, the electrons propagate along chiral quantum Hall edge channels (blue line). (b) Drain lead current of the sample displayed in Fig.~\ref{fig:sample} in a strong magnetic field, as a function of the voltages applied to emitter and detector quantum dot (note that emitter energies increase towards the bottom and detector energies increase towards the left). Depending on the detector quantum dot voltage, the energy introduced by electrons from the source electrode gives rise to either an electron current (red) or a hole current  (blue) through the detector into the drain. Currents are also measured when the detector energy exceeds the emitter energy (relative to the reservoir chemical potential, triangles \textbf{\rom{3}} and \textbf{\rom{4}}).
    }
\end{figure*}

To study relaxation properties of quantum Hall edge channels,~\textcite{lesueur1} devised an experiment in which a quantum point contact induces a non-equilibrium distribution into the outermost of two edge channels, which is energetically probed downstream by a quantum dot. During propagation, interactions between the outer and inner channel cause relaxation of the initial distribution. 
 Surprisingly, a significant amount of energy induced into the setup is lost to inaccessible degrees of freedom.\cite{PhysRevB.81.121302,PhysRevB.81.041311,PhysRevB.84.085105} A further experiment~\cite{Bocquillon2013} to study relaxation in edge channels, that controls the energy of quasi-particle excitations by an RF-circuit,  probes intermediate relaxation by means of a downstream Ohmic contact. The modes of the channels are found to be dissipative, while also in this setup the degrees of freedom that absorb the dissipated energy remain undetermined.

At the ETH Z\"urich an electron spectrometer was used to probe energetic relaxation in  edge channels by means of two successive quantum dots in a novel experiment to identify relaxation mechanisms in quantum Hall edges;~\cite{kraehenmann1,arXivExp} A similar setup has also been realized with shorter propagation paths.\cite{arXivRoche}  
A typical sample, as depicted in Fig.~\ref{fig:sample}, is composed of a source and a drain lead, coupled to an intermediate reservoir via an emitter and a detector quantum dot. The quantum dots act as energy filters for incoming and outgoing electrons. In a strong external magnetic field, the electrons are confined to chiral quantum Hall edge channels.
The current measured in the drain lead of the spectrometer is displayed in Fig.~\ref{fig:current}, as a function of emitter and detector energy.
In addition to signals that arise when the filter energy of the detector quantum dot is tuned below the energy of the emitter quantum dot (triangles \textbf{\rom{1}} and \textbf{\rom{2}} in Fig.~\ref{fig:current}),\footnote{The positive current along the edge of triangle \textbf{\rom{2}} results from higher lying states in the detector dot.\cite{kraehenmann1}} anomalous signatures appear when the detector energy exceeds the emitter energy (triangles \textbf{\rom{3}} and \textbf{\rom{4}}  in Fig.~\ref{fig:current}).

The central goal of this paper is to demonstrate, from  first principles, that the spectroscopic response in \textbf{\rom{3}} and \textbf{\rom{4}} in the ETH spectrometer detector current, in which the detection energy exceeds the emission energy, is a consequence of interactions between electrons in the spatially separated source lead and reservoir region of the device. 
We show that such interactions cause direct transfer of energy to the sample's reservoir, that stems from Auger-like plasmon recombination processes in the source. These processes are enabled by source electrons recombining with holes that are left behind by electrons emitted into the reservoir. 
Our findings indicate that such processes may very well at least partly account for the unexpected energy loss reported in Refs.~[\onlinecite{lesueur1}] and~[\onlinecite{Bocquillon2013}], which calls for controlling the relevant degrees of freedom in upcoming experimental setups, and for the consideration of such processes in further theoretical studies of energy relaxation.
For our purpose, we need to take tunneling as well as  interactions between the  reservoir and  source channel into account in our analysis. 
In addition, 
interactions within the sample's reservoir are considered, that account for the regular features \textbf{\rom{1}} and \textbf{\rom{2}}. 

Theoretical models to capture mechanisms causing relaxation in edge channels often treat electron--electron interactions as a contact interaction, in a perturbative approach~\cite{PhysRevB.81.041311} or employing bosonization techniques. The inflow of electrons into the edge is described either by
 an initial non-equilibrium distribution in the channels,~\cite{PhysRevLett.113.166403,PhysRevB.84.085105} or by a perturbative~\cite{PhysRevLett.109.106403,PhysRevB.82.041306,PhysRevB.96.075144} treatment of tunneling to non-interacting leads. 
	Some studies have taken into account finite or long range interactions to  describe either intra~\cite{PhysRevB.84.085105,PhysRevB.76.085320}- or interchannel~\cite{PhysRevB.84.085105,PhysRevB.85.075309} relaxation, treating tunneling to leads perturbatively, or by considering an initial non-equilibrium distribution in the channels.

To relate distinct contributions observed in the spectrometer's detector current to specific interaction events and transfer processes, we 
employ a perturbative approach within the Keldysh non-equilibrium Green's functions framework in a Fermionic picture. 
The complexity of the diagrammatic approach, treating both interactions and tunneling simultaneously, rapidly increases with the perturbation order. To retain tractability, 
we consider second order Coulomb interactions, which is the lowest order that captures relaxation. This corresponds to transition amplitudes, and thereby to physical processes, that involve the emission and absorption of one virtual photon. 
We limit our analysis to one channel per edge, considering finite range interactions. This constitutes the minimal model to generate current in all triangles \textbf{\rom{1}} to \textbf{\rom{4}} for second order electron--electron interactions, and to thereby capture the experiment's essential features. Tunneling processes beyond one transmission event per quantum dot\footnote{The term transmission event in this context corresponds to the probability of an electron tunneling through a quantum dot in an otherwise decoupled system.} can be neglected if tunnel coupling to the quantum dots is weak, since these processes are typically suppressed by the ratio of the tunnel coupling strength to the intrinsic energy scales of the setup, such as the injection and detection energies as measured from the Fermi sea, or the energy of virtual photons. For finite range interactions, processes in which two electrons pass one of the quantum dots consecutively are furthermore suppressed due to the Pauli principle, by the ratio of the screening length to the length of the spectrometer.\footnote{Coulomb interactions between electrons on the quantum dots and between electrons on the quantum dots and the channels are not taken into account. Such interactions would further suppress consecutive tunneling events due to Coulomb blockade.} Additional features of the detector current, that require going beyond second order perturbation theory in interactions, are discussed in Section~\ref{sec:discussion}.      

The description of the aforementioned Auger-like recombination processes in the source channel, depicted in Fig.~\ref{fig:Processes}, constitutes our central result.
The diagrams which correspond to these processes in our perturbative approach, shown in Figs.~\ref{fig:3b} and~\ref{fig:3c}, generate current exactly in triangles \textbf{\rom{3}} and \textbf{\rom{4}}, as is apparent from Eqs.~(\ref{eq:3bexplicit}) and~(\ref{eq:3cexplicit}). Figure~\ref{fig:TotalCurrent} furthermore shows the current generated in all triangles \textbf{\rom{1}} to \textbf{\rom{4}}, compare Fig.~\ref{fig:current}. Figure~\ref{fig:dynamics} shows the evolution of the initial edge channel distribution for increasing spatial separation of the quantum dots.

\begin{figure}[t]
{\caption{Sketch of the experimental sample: Source (L), drain (R), and reservoir region (I) channels are held at the chemical potentials $\mu_L$, $\mu_R$, and $\mu_I$, respectively. The source and drain channels are coupled to the reservoir region via the emitter quantum dot at energy $\omega_L$ and the detector quantum dot at energy $\omega_R$, with tunneling coupling strength $\Gamma$. \label{fig:SketchSample}}}
{\includegraphics[width=\textwidth]{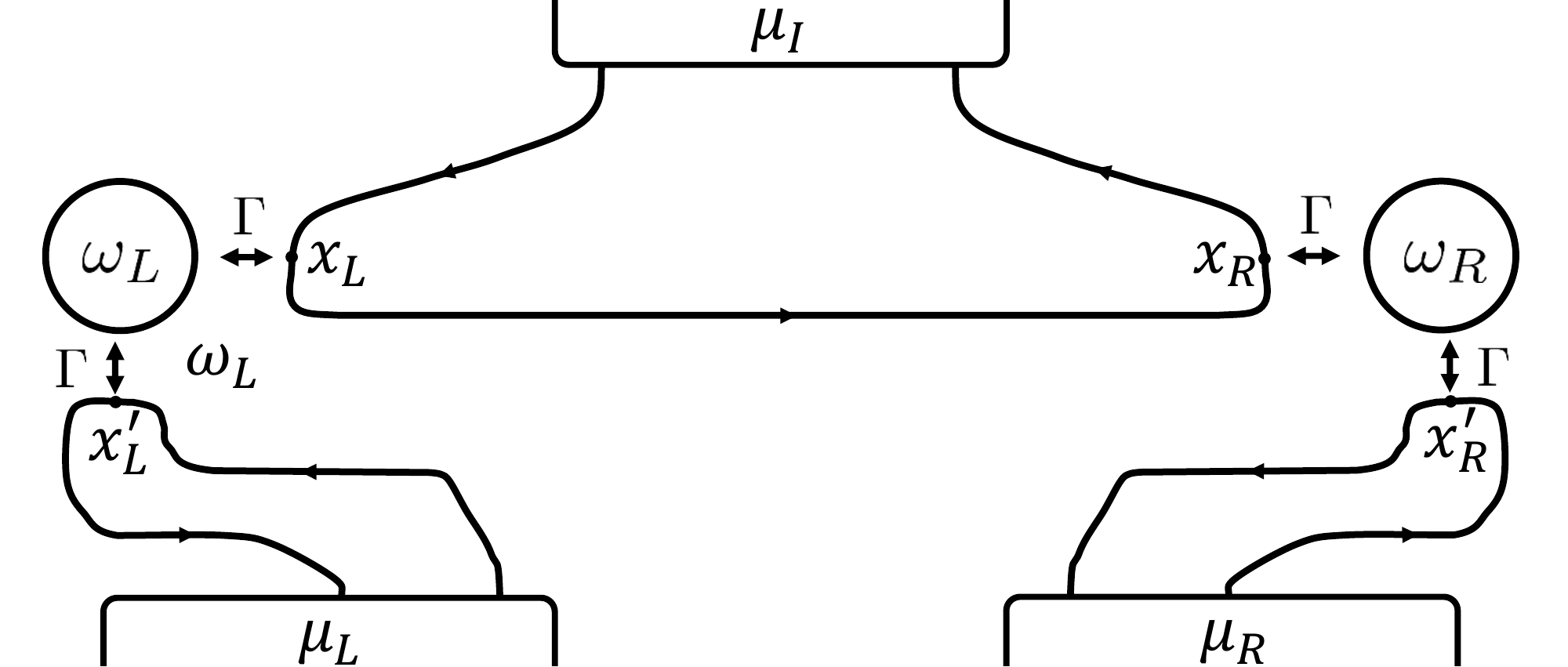}}
\end{figure}

The paper is structured as follows: in Section~\ref{sec:model}, we introduce a model Hamiltonian to describe the ETH sample, devised to capture the distinct features of the current  measured in the detector quantum dot. In Section~\ref{sec:CurrentGeneral} we develop a general expression for the detector current in terms of the Keldysh non-equilibrium Green's functions technique. In Section~\ref{sec:non-interactingI} we solve our model for the case without interactions between charge carriers. In Section~\ref{sec:GFInteracting} we develop systematics in Keldysh space when interactions are taken into account. In Sections~\ref{sec:interactions} and~\ref{sec:HVIL} we introduce interaction terms within the reservoir, as well as between  reservoir and source lead, respectively, before we evaluate the general expressions resulting from our approach for a finite range model interaction in Section~\ref{sec:GaussianModel}.

\section{Model and energy scales}
\label{sec:model}

In this section, we present a model devised to capture the individual contributions \textbf{\rom{1}} to \textbf{\rom{4}}, displayed in Fig.~\ref{fig:current}, to the inelastic current measured in the electron spectrometer. 

A sketch of the sample is shown in Fig.~\ref{fig:SketchSample}: the source channel to the left ($L$), the drain channel to the right ($R$), and the intermediate ($I$) reservoir channel are held at the chemical potentials $\mu_L$, $\mu_R$, and $\mu_I$, respectively. Before interactions are considered, these channels are described as a one-dimensional chiral free electron gas with linear dispersion relation. The spatial coordinate along the channels thereby runs from negative to positive infinity, where  periodic boundary conditions are imposed, and  effects of curvature are not taken into account. Note that we restrict the treatment to a single edge channel per region. The experimental data suggests that interactions are of sufficient range to enable intra-channel relaxation,~\cite{PhysRevB.84.085105,PhysRevB.76.085320} to partially and qualitatively account for the signal in triangles \textbf{\rom{1}} and \textbf{\rom{2}}.

The source lead is tunnel-coupled at position $x_L'$ to the emitter quantum dot, described as a single resonant level $\omega_L$.  
In the same way the drain lead couples  to the detector quantum dot $\omega_R$ at position $x_R'$. Both quantum dots are in turn coupled to the reservoir region at their respective positions $x_L$ and $x_R$ along the reservoir channel. 

While it suffices to take into account interactions within the reservoir channel to generate triangles \textbf{\rom{1}} and \textbf{\rom{2}}, generation of triangles \textbf{\rom{3}} and \textbf{\rom{4}} additionally requires  accounting for interactions between the source and reservoir channels, as will be shown in sections~\ref{sec:interactions} and~\ref{sec:HVIL}. 
The total Hamiltonian of the model  is thus composed of three contributions,
\begin{align}
  H = H_{0} + H_{T} + H_{V}.
\end{align}
$H_0$ contains the description of the individual parts of the system without tunneling or interactions,
 \begin{align}
 \label{eq:H0}
   H_0 = \sum_{\substack{\alpha=L,R \\ k_\alpha }} \epsilon_{k_\alpha} \hat{l}_{k_\alpha}^\dagger \hat{l}_{k_\alpha}^{\vphantom{\dagger}} + \sum_{\alpha=L,R} \omega_\alpha \hat{d}_\alpha^\dagger \hat{d}_\alpha^{\vphantom{\dagger}}
    + \sum_{k} \epsilon_k \hat{r}_k^{\dagger}  \hat{r}^{\vphantom{\dagger}}_k. 
 \end{align}
Here $\hat{l}_{k_\alpha}$  denote the fermionic annihilation operators of the left and right leads, $ \hat{d}_\alpha$ of the left and right quantum dots, and $ \hat{r}^{\vphantom{\dagger}}_k$ of the intermediate reservoir region. The tunneling Hamiltonian
\begin{align}
  H_T &= \sum_{\substack{\alpha=L,R \\ k_\alpha }} \left[ t_{k_\alpha \alpha} \hat{l}_{k_\alpha}^{\dagger} \hat{d}_\alpha^{\vphantom{\dagger}} + t_{k_\alpha \alpha}^*  \hat{d}_\alpha^{\dagger} \hat{l}_{k_\alpha}^{\vphantom{\dagger}}\right] \nonumber \\
  &+  \sum_{\substack{\alpha=L,R \\ k }} \left[ t_{k \alpha} \hat{r}_{k}^{\dagger} \hat{d}_\alpha^{\vphantom{\dagger}} + t_{k \alpha}^*  \hat{d}_\alpha^{\dagger} \hat{r}_{k}^{\vphantom{\dagger}} \right]
\end{align} 
describes tunneling coupling between the lead channels and the dots, as well as between the dots and the reservoir region. The tunneling amplitudes $t_{k_\alpha \alpha} = t \exp (-i k_\alpha x_\alpha')$ and $t_{k \alpha} = t \exp (-i k x_\alpha)$ contain the system's spatial information and describe the local coupling of the leads to the left and right dots at $x_L'$ and $x_R'$, as well as the local coupling of the dots to the reservoir at $x_L$ and $x_R$, respectively, $x_L < x_R$. 
The amplitudes $t$ are assumed to be momentum independent and equal for all tunneling to the left and to the right quantum dot, and determine, in combination with the density of states $\rho$ in the respective lead, the tunneling coupling strength $\Gamma = 2\pi |t|^2 \rho$.
The interaction Hamiltonian
\begin{align}
  H_V = H_V\left( \left\{ \hat{l}_{k_\alpha}^{\vphantom{\dagger}}, \hat{l}_{k_\alpha}^{\dagger},  \hat{r}_k^{\vphantom{\dagger}}, \hat{r}_k^{\dagger} \right\} \right)
\end{align}
generally depends on the lead and reservoir creation and annihilation operators in a way that will be specified when the respective interactions are accounted for in Sections~\ref{sec:interactions} and~\ref{sec:HVIL}.

Energy scales in the experiment are approximately $\Gamma \sim 1\mu$eV $< k_B T \sim 2.5\mu$eV  $ < \hbar v/ \Delta x \sim 30\mu$eV $ 
 < \mu_L - \mu_R \sim 400\mu$eV,\cite{kraehenmann1} where $T$ is the temperature of the sample. 
Due to the proximity of the top gates to the two-dimensional electron gas in the experiment ($\sim\!\!90$nm), as our main assumption, we estimate the screening length $\lambda$ to be by an order of magnitude shorter than the distance between the quantum dots, $\lambda \ll \Delta x$. 
Calculations are performed at $T=0$, which gives rise to an error whenever $\omega_L$ and $\omega_R$ are within a range of $k_B T$ to either chemical potential $\mu_L$ or $\mu_R$ for elastic transfer, or within a range of $k_B T$ to the outlines of triangles \textbf{\rom{1}} to \textbf{\rom{4}} for inelastic transfer, as is illustrated in Section~\ref{sec:1a}.

\section{General expression for drain lead current}
\label{sec:CurrentGeneral}

The central observable in the electron spectrometer is the current through the drain lead. In this section, a general expression for this current is stated, which serves as the basis for all further considerations. Within the framework of the Keldysh non-equilibrium Green's function formalism, the goal is to express the current in terms of the full Green's functions of the reservoir region, containing both information about tunneling as well as about interactions, following the approach of Ref.~[\onlinecite{meir1}].

The current through the drain lead is given by
\begin{align}
  I_R &= - e \left< \dot{n}_R(t) \right> \nonumber \\
         &= -\frac{ie}{\hbar} \sum_{k_R} \left[ t_{k_R R} \left<\hat l_{k_R}^\dagger(t) d_R^{\vphantom{\dagger}}(t)\right> - t_{k_R R}^*  \left< d_R^{\dagger}(t) \hat l_{k_R}^{\vphantom{\dagger}}(t) \right> \right].
\end{align}
After Fourier transformation, we obtain
\begin{align}
\label{eq:currentinitial}
I_R = -\frac{e}{2\pi} \int_{-\infty}^{\infty} d\omega \sum_{k_R} \left[ t_{k_R R} G_{R k_R}^<(\omega) - t_{k_R R}^* G_{k_R R}^<(\omega)  \right],
\end{align}
where we used the definition of the lesser Green's function $G_{R k_R}^<(t,t) = i\left<\hat l_{k_R}^\dagger(t) d_R^{\vphantom{\dagger}}(t)\right>\big/\hbar$ on the Keldysh contour.

In the following we set $\hbar = k_B = 1$ and adopt a matrix notation in which $\text{D}_{\text{L}/\text{R}}$ denotes the index of the left/right quantum dot, $\text{L}_{\text{L}/\text{R}}$ denotes the indices of the left/right lead, and I denotes the indices of the intermediate reservoir region. In this section, all Green's functions depend on the energy $\omega$, which allows us to suppress this dependency in our notation. Thus, (\ref{eq:currentinitial}) is written as
\begin{align}
\label{eq:currentinitial2}
&I_R = \nonumber \\ &-\frac{e}{2 \pi} \int_{-\infty}^{\infty} d\omega \, \text{tr} \left\{ \mathbf{t}_{\text{L}_\text{R} \text{D}_\text{R}} \mathbf{G}^<_{\text{D}_\text{R} \text{L}_\text{R}}   - \mathbf{G}^<_{\text{L}_\text{R} \text{D}_\text{R}} \mathbf{t}_{\text{D}_\text{R} \text{L}_\text{R}}    \right\}.
\end{align}
Taking the detector quantum dot and the right lead to be non-interacting allows to recast~(\ref{eq:currentinitial2}) into the desired general form
\begin{align}
\label{eq:currentfinal}
&I_R = \nonumber \\ &-\frac{e}{2 \pi} \int_{-\infty}^{\infty} d\omega \, \text{tr} \big\{ \mathbf{\overline{g}}^a_{\text{D}_\text{R} \text{D}_\text{R}} \mathbf{\Sigma}^<_{\text{D}_\text{R} \text{L}_{\text{R}} \text{D}_\text{R}} \mathbf{\overline{g}}^r_{\text{D}_\text{R} \text{D}_\text{R}} \mathbf{\Sigma}^>_{\text{D}_\text{R} \text{I} \text{D}_\text{R}} \nonumber \\   
& \qquad \qquad \qquad -\mathbf{\overline{g}}^a_{\text{D}_\text{R} \text{D}_\text{R}} \mathbf{\Sigma}^>_{\text{D}_\text{R} \text{L}_{\text{R}} \text{D}_\text{R}} \mathbf{\overline{g}}^r_{\text{D}_\text{R} \text{D}_\text{R}} \mathbf{\Sigma}^<_{\text{D}_\text{R} \text{I} \text{D}_\text{R}}    \big\},
\end{align}
as is shown in Appendix~\ref{app:DrainCurrent}. In~(\ref{eq:currentfinal}),
\begin{align}
\label{eq:gdotleadR}
  \mathbf{\overline{g}}_{\text{D}_\text{R} \text{D}_\text{R}} = \mathbf{g}_{\text{D}_\text{R} \text{D}_\text{R}} + \mathbf{g}_{\text{D}_\text{R} \text{D}_\text{R}} \mathbf{\Sigma}_{\text{D}_\text{R} \text{L}_\text{R} \text{D}_\text{R}}  \mathbf{\overline{g}}_{\text{D}_\text{R} \text{D}_\text{R}},
\end{align}
describes charge carriers passing back and forth between the detector dot and the right lead, where
\begin{align}
\label{eq:selfenergylead}
  \mathbf{\Sigma}_{\text{D}_\text{R} \text{L}_\text{R} \text{D}_\text{R}} =  \mathbf{t}_{\text{D}_\text{R} \text{L}_\text{R}} \mathbf{g}_{\text{L}_\text{R} \text{L}_\text{R}} \mathbf{t}_{\text{L}_\text{R} \text{D}_\text{R}}
\end{align}
denotes the tunneling self-energy of the isolated drain lead. The tunneling self-energy of the interacting reservoir region
\begin{align}
\label{eq:tunnelingselfenergyreservoir}
  \mathbf{\Sigma}_{\text{D}_\text{R} \text{I} \text{D}_\text{R}} =  \mathbf{t}_{\text{D}_\text{R} \text{I}} \mathbf{G}_{\text{I} \text{I}} \mathbf{t}_{\text{I} \text{D}_\text{R}}
\end{align}
contains the Green's function of the reservoir region $\mathbf{G}_{\text{II}}$ which develops with the total Hamiltonian $H$. In the following sections, this Green's function will be evaluated for different types of interactions, and determines the drain current.

Written explicitly, the general expression for the current~(\ref{eq:currentfinal}) reads
\begin{align}
\label{eq:currentfinalexplicit}
I_R &= -\frac{ie}{2\pi} \int_{-\infty}^{\infty} d\omega A_{\text{D}_{\text{R}}} (\omega) \sum_{k,k'} \big[ f_R(\omega) t_{R k} G^>_{kk'} (\omega) t_{k' R} \nonumber \\ 
   &\qquad \qquad \qquad- \left(f_R(\omega) -1 \right) t_{R k} G^<_{kk'}(\omega) t_{k' R} \big].
\end{align}
Here
\begin{align}
  A_{\text{D}_{\text{R}}} (\omega) = \frac{\Gamma(\omega)}{(\omega-\omega_R)^2 + \frac{\Gamma(\omega)^2}{4}}
\end{align}
reflects the broadening of the detector quantum dot's energy level due to the coupling to the drain lead.\footnote{Taking into account also the broadening by the coupling to the reservoir generates the transition coefficient~(\ref{eq:transition}). } The occupation of the drain lead is described by the Fermi distribution  $f_R$, and the tunneling coupling strength $\Gamma(\omega)$ is here determined by the density of states $\rho(\omega)$ in the drain lead.

\section{Non-interacting case}
\label{sec:non-interactingI}

 \begin{figure*}[t!]
  \subfloat[$\mu_I>\mu_R$\label{fig:elastic1}]{%
    \includegraphics[height=.3\textwidth]{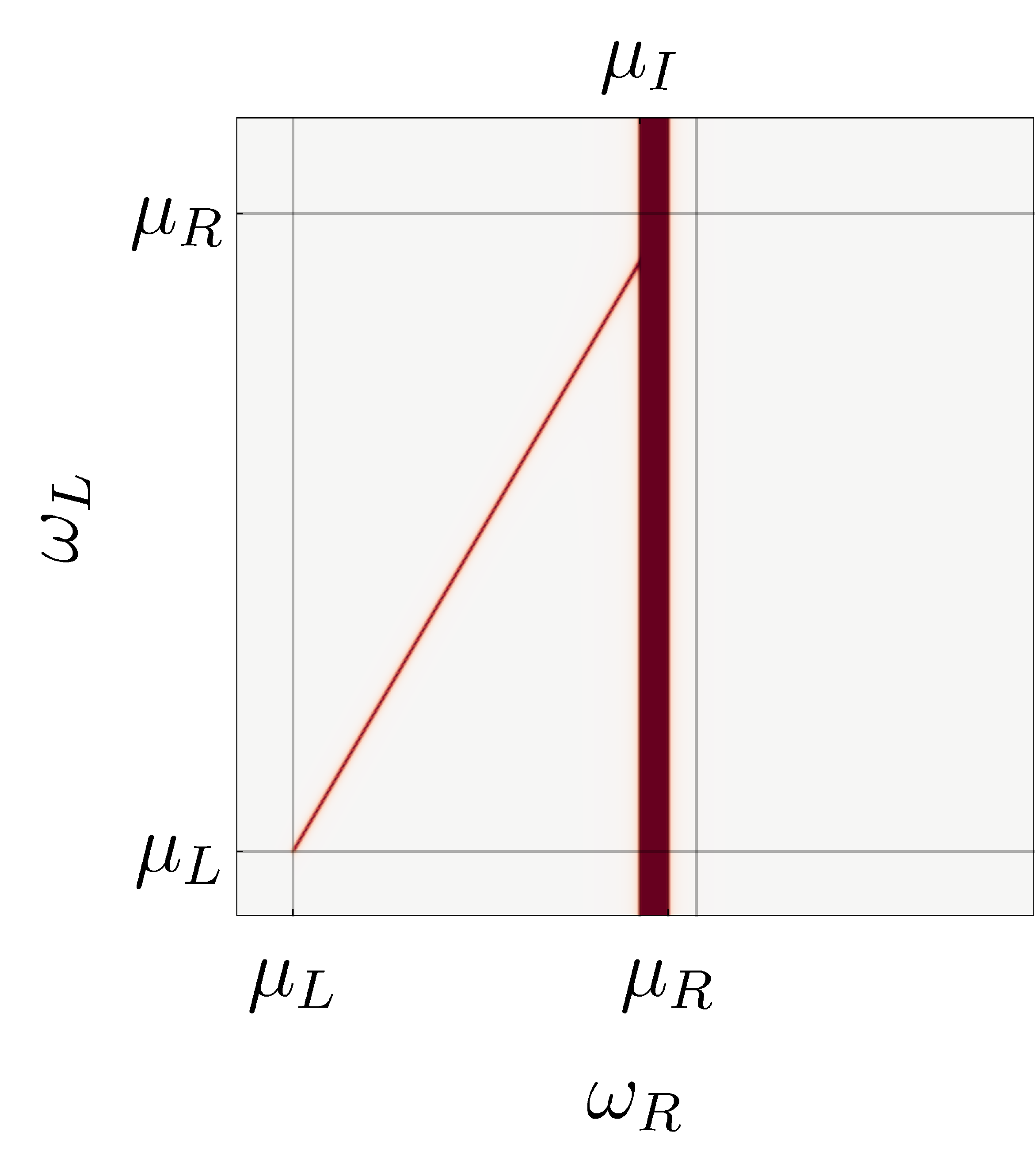}
  }
  \hfill
      \subfloat[$\mu_I=\mu_R$\label{fig:elastic2}]{%
    \includegraphics[height=.3\textwidth]{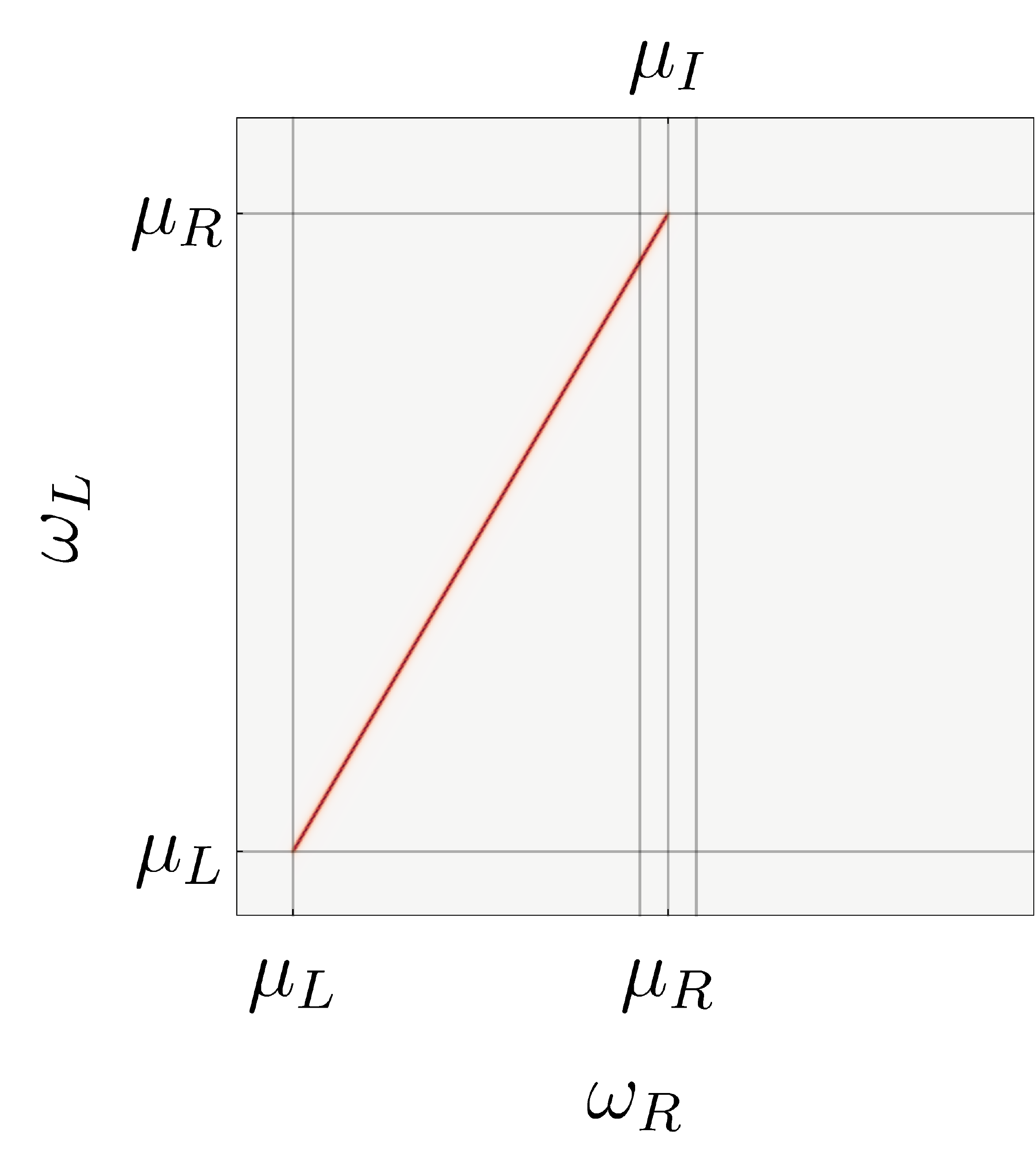}
  }
  \hfill
      \subfloat[$\mu_I<\mu_R$\label{fig:elastic3}]{%
    \includegraphics[height=.3\textwidth]{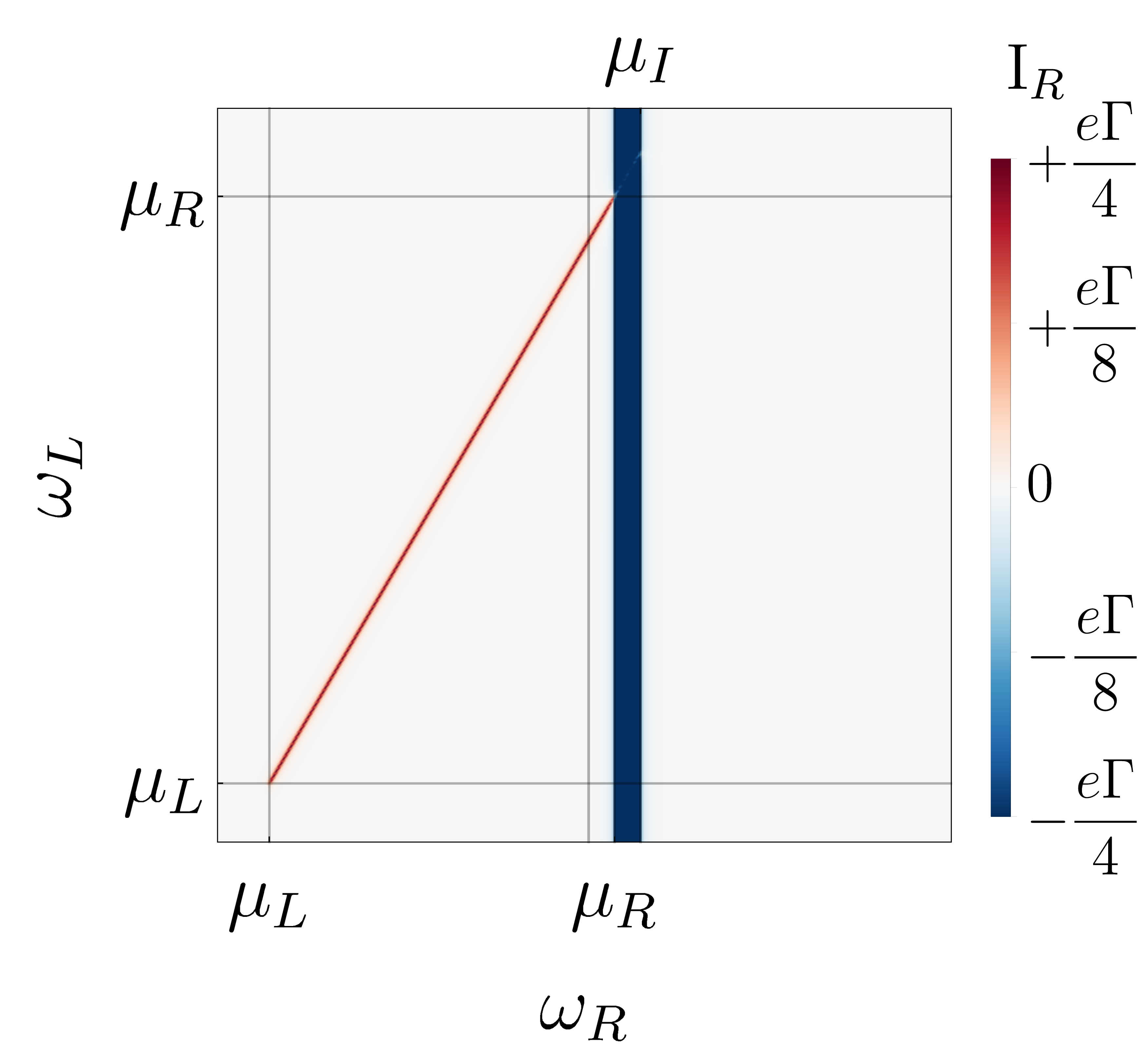}
  }  
    \caption{Current~(\ref{eq:Inonint}) through drain in absence of interactions, for large bias between source and drain channels, and for~(a) slightly positive,~(b) zero, and~(c) slightly negative bias between reservoir and drain channel. Note that dot energies increase from top to bottom and from right to left and that the aspect ratio has been adjusted, so as to match the conventions used in the experimental data shown in Fig.~\ref{fig:current}. (Parameters are $\mu_L = \mu_R + 400 \Gamma$, $\mu_I =$ (a): $\mu_R + 30\Gamma$, (b): $\mu_R$, (c): $\mu_R - 30\Gamma$.)  }
  \label{fig:elasticcurrent}
\end{figure*}

In this section, we solve the non-interacting case, setting $H_V \equiv 0$. We treat the intermediate reservoir region as a one-dimensional chiral channel with right-moving particles and linear dispersion relation, such that $\epsilon_{k} = vk$ in~(\ref{eq:H0}), where $v$ is the channel's Fermi velocity.

In this case, the Green's function of the reservoir region develops according to the Dyson equation
\begin{align}
\label{eq:tunnelingG}
  \mathbf{G}^T_{\text{II}} = \mathbf{g}_{\text{II}} + \mathbf{g}_{\text{II}} \mathbf{\Sigma}_\text{ITI} \mathbf{G}^T_{\text{II}},
\end{align}
with the tunneling self-energy
\begin{align}
  \mathbf{\Sigma}_\text{ITI} = \mathbf{\Sigma}_{\text{I} \text{D}_\text{L} \text{I} } + \mathbf{\Sigma}_{\text{I} \text{D}_\text{R} \text{I} },
\end{align}
in which
\begin{align}
\mathbf{\Sigma}_{\text{I}\text{D}_{\text{L/R}}\text{I}} &= \mathbf{t}_{\text{I} \text{D}_\text{L/R}} \mathbf{\overline{g}}_{\text{D}_\text{L/R} \text{D}_\text{L/R}} \mathbf{t}_{\text{D}_\text{L/R} \text{I}}.
\end{align}
The superscript $T$ in~(\ref{eq:tunnelingG}) indicates that these Green's functions of the intermediate region develop without interaction terms.

The kinetic equation for the lesser component of~(\ref{eq:tunnelingG}) reads
\begin{align}
\label{eq:GTkinetic}
  \mathbf{G}^{T<}_{\text{II}} = &\left(\mathbf{I} + \mathbf{G}^{Tr}_{\text{II}}  \mathbf{\Sigma}^r_\text{ITI} \right) \mathbf{g}^<_{\text{II}} \left(\mathbf{I} + \mathbf{\Sigma}^a_\text{ITI} \mathbf{G}^{Ta}_{\text{II}}  \right) \nonumber \\ 
  &+ \mathbf{G}^{Tr}_{\text{II}} \mathbf{\Sigma}^<_\text{ITI} \mathbf{G}^{Ta}_{\text{II}},
\end{align}
with ``$< \rightarrow >$'' for the greater component. 
Here, the superscripts $r$ and $a$ denote the retarded and advanced components of the respective Green's functions. 
Upon insertion into~(\ref{eq:currentfinal}), the first term of~(\ref{eq:GTkinetic}) gives rise to a current exchanged directly between the intermediate reservoir and the drain lead which depends on the respective chemical potentials of those regions $\mu_I$ and $\mu_R$. The second term of~(\ref{eq:GTkinetic}) gives rise to a current depending on the chemical potentials $\mu_L$ and $\mu_R$ of the source and the drain lead. Evaluation of the respective transition self-energies is presented in Appendix~\ref{app:CurrentNoV}.

The explicit expression for the current of the non-interacting system then reads
\begin{align}
\label{eq:currentexplicit}
  I_R = &\frac{e}{2\pi} \int_{-\infty}^{\infty} d\omega \left[ f_I(\omega) - f_R(\omega) \right] \mathcal{R}_L (\omega)  \mathcal{T}_R (\omega) \nonumber \\
   +&\frac{e}{2\pi} \int_{-\infty}^{\infty} d\omega \left[ f_L(\omega) - f_R(\omega) \right] \mathcal{T}_L (\omega)  \mathcal{T}_R (\omega),
\end{align}
with  transmission coefficient 
\begin{align}
\label{eq:transition}
  \mathcal{T}_{L/R} (\omega) = \frac{\Gamma(\omega)^2}{(\omega-\omega_{L/R})^2 + \Gamma(\omega)^2}
\end{align}
and  reflection coefficient
\begin{align}
  \mathcal{R}_{L/R} (\omega) = 1 - \mathcal{T}_{L/R} (\omega).
\end{align}
Since no interactions between electrons have been considered so far, the explicit formula for the current through the drain lead~(\ref{eq:currentexplicit}) coincides with the Landauer-B\"uttiker result.\cite{meir1} The first line in~(\ref{eq:currentexplicit}), which describes current  exchanged directly between the reservoir region and the drain, shows that charge carriers provided by the reservoir must be reflected at the emitter quantum dot before they can be transmitted through the detector, see Fig.~\ref{fig:SketchSample}. The second line in~(\ref{eq:currentexplicit}) requires transmission through both quantum dots at the same energy, and therefore constitutes the contribution of elastically transferred electrons.

\begin{figure*}[t]
  \subfloat[Tunneling-dressed polarization diagram\label{fig:PolGen}]{%
    \includegraphics[width=.48\textwidth]{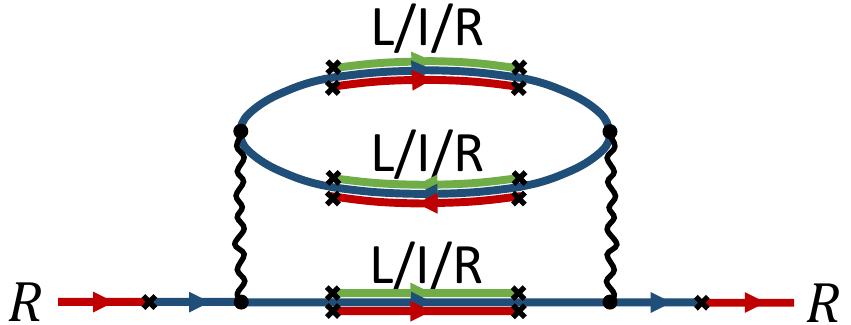}
  }
  \hfill
      \subfloat[Tunneling-dressed exchange diagram\label{fig:ExchGen}]{%
    \includegraphics[width=.48\textwidth]{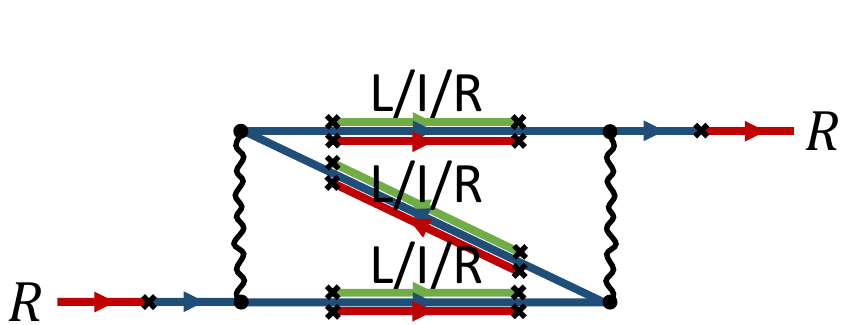}
  }
  
    \caption{Tunneling-dressed (a) polarization  and (b) exchange  self-energy diagrams. Diagrams that determine the current through the drain are terminated by Green's functions of this channel (red lines). Tunneling (black crosses) from the intermediate region (blues lines) to the source (red lines) and drain channel through emitter and detector quantum dot, respectively, generates 27 diagrams for both self-energies. These 27 diagrams correspond to 27 distinct  physical processes. Limiting tunneling as described in Section~\ref{sec:interactions}, only three of those processes remain.
      }
  \label{fig:ExchPolGeneral}
\end{figure*}

At zero temperature and for constant density of states in the source, drain, and reservoir region, i.e.~constant $\Gamma$, (\ref{eq:currentexplicit}) is given by
\begin{align}
 \label{eq:Inonint}
  I_R = \frac{e\Gamma}{2\pi} &\left[ \arg\left( \omega_R-\mu_I+i \Gamma \right) - \arg\left( \omega_R-\mu_R+i \Gamma \right) \right] \nonumber \\
   - &I_{\text{el}} (\mu_I, \mu_R) +  I_{\text{el}} (\mu_L, \mu_R),
\end{align}
where for the elastic current transmitted through both dots, which corresponds to the second line of~(\ref{eq:currentexplicit}),  we have 
\begin{align}
 \label{eq:Ielastic}
& I_{\text{el}}(\mu_L,\mu_R) =   \frac{e\Gamma}{2\pi} \frac{\Gamma^2}{(\omega_L-\omega_R)^2 + (2\Gamma)^2} \nonumber \\
  &\times \Bigg\{ \arg\left( \omega_L-\mu_L+i \Gamma \right) - \arg\left( \omega_L-\mu_R+i \Gamma \right) \nonumber \\
  &\quad+\arg\left( \omega_R-\mu_L+i \Gamma \right) - \arg\left( \omega_R-\mu_R+i \Gamma \right) \nonumber \\
  &\quad+\frac{\Gamma}{\omega_L-\omega_R} \nonumber \\
   &\quad\times\log\left[\frac{\left( (\omega_L-\mu_R)^2 +\Gamma^2 \right)\left( (\omega_R-\mu_L)^2 +\Gamma^2 \right)}{\left( (\omega_L-\mu_L)^2 +\Gamma^2 \right)\left( (\omega_R-\mu_R)^2 +\Gamma^2 \right)}\right] \Bigg\}.
\end{align}
The current~(\ref{eq:Inonint}) is displayed in Fig.~\ref{fig:elasticcurrent}, for large bias between source and drain lead, and for~(\ref{fig:elastic1}) slightly positive,~(\ref{fig:elastic2}) zero, and~(\ref{fig:elastic3}) slightly negative bias between reservoir and drain lead.

\section{Green's functions in the presence of interactions}
\label{sec:GFInteracting}

In this section, a systematic scheme to develop the contour ordered Green's functions is presented, 
for the case in which interactions within the reservoir region are considered.

The central object to determine the current through the drain lead~(\ref{eq:currentfinal},\ref{eq:currentfinalexplicit}) are the lesser and greater Green's functions of the reservoir region, governed by the total Hamiltonian $H$, for which we have the kinetic equation
\begin{align}
\label{eq:Gkinetic}
  \mathbf{G}^{<}_{\text{II}} = &\left(\mathbf{I} + \mathbf{G}^{r}_{\text{II}}  \mathbf{\Sigma}^r_\text{II} \right) \mathbf{g}^<_{\text{II}} \left(\mathbf{I} + \mathbf{\Sigma}^a_\text{II} \mathbf{G}^{a}_{\text{II}}  \right) \nonumber \\ 
  &+ \mathbf{G}^{r}_{\text{II}} \mathbf{\Sigma}^<_\text{II} \mathbf{G}^{a}_{\text{II}},
\end{align}
with the replacement ``$< \rightarrow >$'' for the greater component. Since Green's functions and self-energies in~(\ref{eq:Gkinetic}) depend both on tunneling and interaction Hamiltonians, $H_T$ and $H_V$, respectively, a method to separate those dependencies is desirable. 

In a diagrammatic expansion of Green's functions which include both these types of processes, it is possible to rearrange terms such that each interaction event is succeeded by all possible tunneling events, before the next interaction event is considered.
This leads to the Dyson equation
\begin{align}
\label{eq:altG}
  \mathbf{G}_{\text{II}} = \mathbf{G}^{T}_{\text{II}} + \mathbf{G}^{T}_{\text{II}} \mathbf{\Sigma}_\text{IVI} \mathbf{G}_{\text{II}}
\end{align}
 for the Green's function of the reservoir region. Here, the tunneling Green's function $\mathbf{G}^{T}_{\text{II}}$, previously encountered in~(\ref{eq:tunnelingG}), develops only with the tunneling self-energy, and all bare Green's function lines $\mathbf{g}_{\text{II}}$ in the usual interacting self-energy without tunneling are replaced by $\mathbf{G}^{T}_{\text{II}}$. This prescription defines the self-energy $\mathbf{\Sigma}_\text{IVI}$.

In a next step, the Dyson equation~(\ref{eq:altG}) is truncated after the first iteration, i.e.
\begin{align}
\label{eq:altGtruncated}
  \mathbf{G}_{\text{II}} \simeq \mathbf{G}^{T}_{\text{II}} + \mathbf{G}^{T}_{\text{II}} \mathbf{\Sigma}_\text{IVI} \mathbf{G}^{T}_{\text{II}},
\end{align}
which  allows us to take into account the second order of interactions
in the interacting self-energy $\mathbf{\Sigma}_\text{IVI}$ in section~\ref{sec:interactions}. Applying the Langreth rules to~(\ref{eq:altGtruncated}), we find
\begin{align}
\label{eq:altGLangreth}
  &\mathbf{G}^<_{\text{II}} \simeq \mathbf{G}^{T<}_{\text{II}} + \mathbf{G}^{T<}_{\text{II}} \mathbf{\Sigma}^a_{\text{IVI}} \mathbf{G}^{Ta}_{\text{II}} \nonumber \\ 
  &\qquad\quad+ \mathbf{G}^{Tr}_{\text{II}} \mathbf{\Sigma}^<_{\text{IVI}} \mathbf{G}^{Ta}_{\text{II}} + \mathbf{G}^{Tr}_{\text{II}} \mathbf{\Sigma}^r_{\text{IVI}} \mathbf{G}^{T<}_{\text{II}},
\end{align}
with ``$< \rightarrow >$'' for the greater component of the Green's function. The first term in~(\ref{eq:altGLangreth}) gives rise to the non-interacting current of section~\ref{sec:non-interactingI}.  The second and fourth terms in~(\ref{eq:altGLangreth}) constitute corrections to this current due to the interaction term. This follows from the fact that the lesser tunneling Green's functions (and the greater tunneling Green's functions of the corresponding expression) in~(\ref{eq:altGLangreth}) are evaluated at the same energy $\omega$ as the lesser and greater Green's function of the drain lead, which enter the general formula of the current~(\ref{eq:currentfinal}) via the corresponding tunneling self-energies, $\mathbf{\Sigma}^{</>}_{\text{D}_\text{R} \text{L}_{\text{R}} \text{D}_\text{R}}$. Thereby, the current generated by the latter terms is confined to the same regions in the $\omega_L/\omega_R$ space as the elastic current shown in Fig.~\ref{fig:elasticcurrent}. 

The third term in~(\ref{eq:altGLangreth}) gives rise to inelastic contributions to the current, since the lesser/greater interacting self-energies entail equilibration at energies different from $\omega$, as is demonstrated in the following section.

\section{Interactions in the reservoir region}
\label{sec:interactions}

In this section, Coulomb interaction between electrons in the reservoir region is explicitly taken into account. The interaction is described by the Hamiltonian
\begin{align}
\label{eq:HV}
  H_V^{I} = \frac{1}{2\Omega} \sum_{k,k',q} \nu_q^{\vphantom{\dagger}} \hat{r}^\dagger_{k-q} \hat{r}^\dagger_{k'+q} \hat{r}_{k'}^{\vphantom{\dagger}} \hat{r}_{k}^{\vphantom{\dagger}},
\end{align}
where
\begin{align}
\label{eq:nuq}
  \nu_q^{\vphantom{\dagger}} = \int_{-\infty}^{\infty} \, dr \gamma_r e^{i q r}
\end{align}
is the Fourier transform of the Coulomb matrix element $\gamma_r$ in real space. 
In a diagrammatic approach, the first order Hartree and Fock contributions contain divergent terms, which have been shown via a bosonization calculation to correspond to a mere shift of the chemical potential.\cite{Neuenhahn2008} The converging part of the exchange contribution additionally gives rise to a momentum dependent renormalization of velocities, which saturates above $1/\lambda$. Below this value, second order diagrams largely cancel. In Keldysh space, the first order contributions act in the same way as one-body potentials, and therefore do not give rise to inelastic processes.\footnote{In Keldysh space, one-body potentials only affect retarded and advanced self-energies. The greater and lesser components of the self-energy are unaffected.\cite{Maciejko2007} Only the latter cause inelastic currents. }
Thus, we  neglect these contributions, and exclusively consider the polarization and exchange diagrams in the interacting self-energy, which are the lowest order diagrams generating inelastic  current. 
The tunneling-dressed versions of these diagrams, which correspond to the third term of~(\ref{eq:altGLangreth}), are displayed in Fig.~\ref{fig:ExchPolGeneral}.

In the following, we consider equal chemical potentials in the reservoir and drain lead regions, $\mu_I=\mu_R$. 
We consider one transmission event per quantum dot, i.e.~we collect all tunneling amplitudes which combine to the product $\mathcal{T}_L \mathcal{T}_R$, cf.~(\ref{eq:transition}), discarding further tunneling processes. In this way, we disregard processes in which more than one electron passes through the same quantum dot. 
 For contact interactions, such processes are precluded by the Pauli exclusion principle, since in this case the emitted electron and excited electrons arrive at the detector at the same time. 
Explicit evaluation of such processes for finite-range interactions shows that the approximation is justified as long as the screening length is smaller than the dot separation, $\lambda \ll \Delta x$.
The approximation overestimates currents within a range of $\Gamma$ 
of the lines of zero-momentum transfer, i.e.~the hypotenuses of triangles \textbf{\rom{1}} to \textbf{\rom{4}}, for inelastic transfer.  
Furthermore, processes in which the initial source electron returns into the emitter quantum dot after interacting with electrons in the reservoir are neglected. 

\subsection{Corrections to elastic current}
\label{sec:CorrElastic}

\begin{figure}[t]
    \includegraphics[width=.82\textwidth]{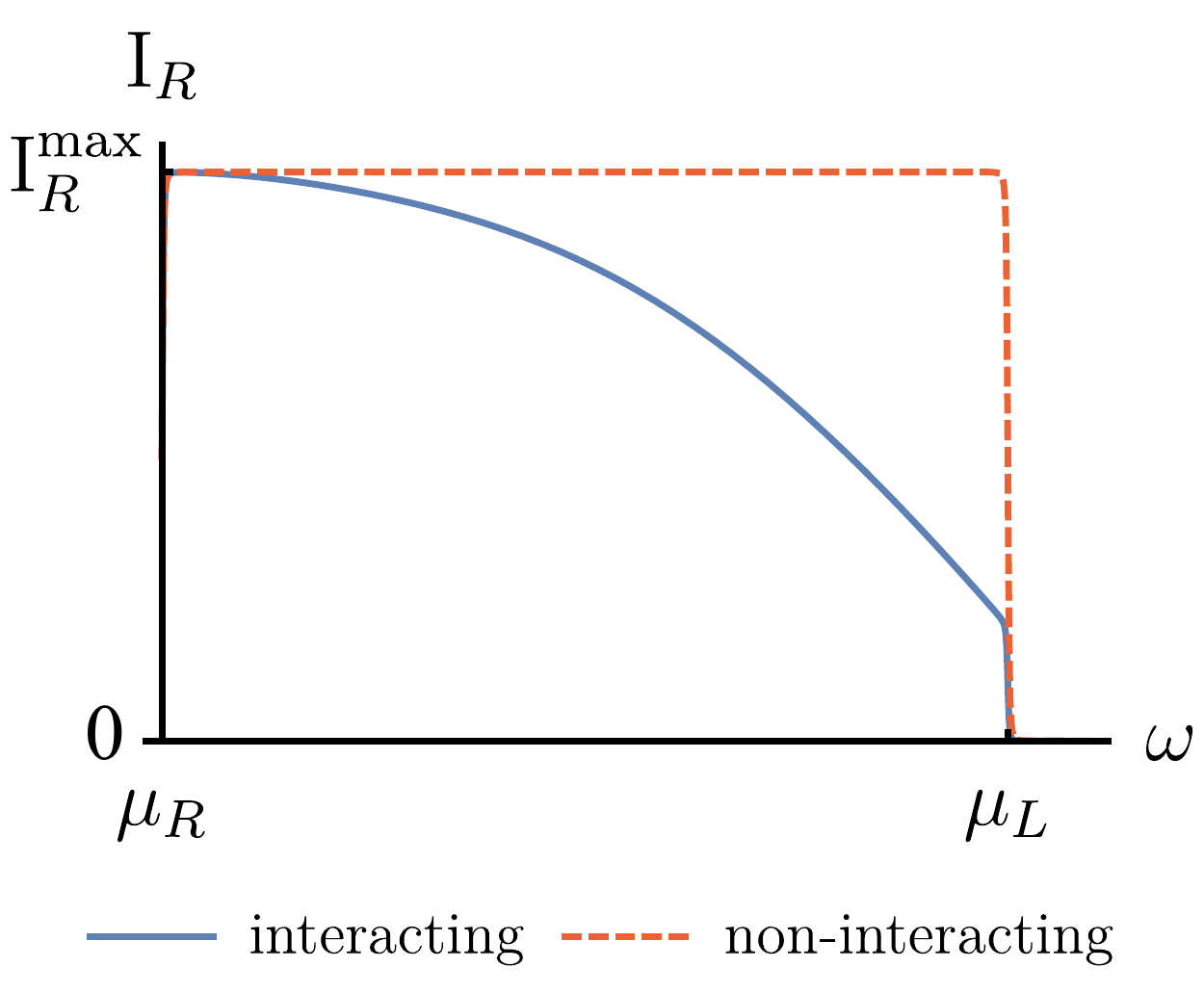}
    \caption{Corrections due to interactions between electrons  
    cause the elastic current~(\ref{eq:correlastic}) to decrease from the maximum value $\text{I}_R^{\text{max}}$ as the distance $\omega$ = $\omega_R$ = $\omega_L$ of emitter and detector quantum dot energies  is increased from the Fermi level $\mu_R$. Elastic current close to zero 
     indicates parameter values for which the perturbative approach to second order in interactions becomes unphysical, cf~(\ref{eq:correlastic}) and (\ref{eq:validity}). 
    (Plot includes~(\ref{eq:Epol}), (\ref{eq:Eexch}), and interchannel  correction~(\ref{eq:EpolIL}), parameters are $\mu_L = \mu_R + 400\Gamma$, $v = 260 \lambda \Gamma$, $\nu_0 = 640\lambda \Gamma$, $\Delta x = 8\lambda$, $d =2.8\lambda$.)}
  \label{fig:ElasticCorrectedI}
\end{figure}

The interaction corrections to the elastic current are the second and fourth term in expansion~(\ref{eq:altGLangreth}). The explicit expression for the correction in terms of Green's functions is contained in Appendix~\ref{app:ElasticCorrection}.

After collection of those tunneling amplitudes which combine to the product of the transition probability through the left and the right dot~(\ref{eq:transition}), evaluation of the correction terms requires the retarded and advanced components of the self-energy $ \mathbf{\Sigma}_{\text{IVI}}$ to zeroth order in tunneling amplitudes, denoted by $\Sigma$. These components are obtained from the respective lesser/greater elements\footnote{the lesser and greater components on the Keldysh contour follow from the Langreth rules.~\cite{haug1}} upon employing the Kramers-Kronig relation
\begin{align}
\label{eq:KK}
  \mathbf{G}^{r/a}(\omega) = \frac{i}{2\pi} \int_{-\infty}^{\infty} d\omega' \frac{\mathbf{G}^{>}(\omega')-\mathbf{G}^{<}(\omega')}{\omega-\omega'\pm i\delta}.
\end{align}
The matrix elements are given by
\begin{align}
\label{eq:sigma}
  \Sigma_k(\omega)^{r/a} = g_k^{r/a}(\omega)E^2(vk). 
\end{align}
Here we have defined $E^2 = E^2_{\text{pol}}+E^2_{\text{exch}}$ to which, at zero temperature, the polarization diagram contributes (see also Ref.~[\onlinecite{Neuenhahn2008}]) with
\begin{align}
\label{eq:Epol}
   E^2_{\text{pol}}(\omega) = \frac{1}{(2\pi v)^2} \int_{0}^{\mu_I -\omega} \! \!\!\!\!\!\!\! d\omega'\, \omega'  \nu_{\omega'/v}^2,
\end{align}
and the exchange diagram contributes
\begin{align}
\label{eq:Eexch}
   E^2_{\text{exch}}(\omega) = \frac{1}{(2\pi v)^2} \int_{0}^{\mu_I-\omega} \! \!\!\!\!\!\!\!  d\omega''\, \nu_{\omega''/v} \int_{\mu_I -\omega-\omega''}^{\mu_I -\omega} \!\!\!\! \! \!\!\!\!\!\!\! d\omega'\,   \nu_{\omega'/v}.
\end{align}
For $E^2$ decaying sufficiently fast as $\omega \to \infty$ in the complex plane, and upon neglecting poles of $E^2$, which holds for $\lambda \ll x$, the corrected elastic current becomes
\begin{align}
\label{eq:correlastic}
  I_R^{\text{el,corr}} = \frac{e}{2\pi} \int_{-\infty}^{\infty} d\omega \left[ f_L(\omega) - f_R(\omega) \right] \mathcal{T}_L (\omega)  \mathcal{T}_R (\omega) \nonumber\\
  \times\left[ 1- \frac{\Delta x^2}{v^2} E^2(\omega) + \frac{\partial^2}{\partial \omega^2} E^2(\omega) \right],
\end{align}
cf.~Appendix~\ref{app:ElasticCorrection}. The corrected current changes direction when the bracket in the second line of~(\ref{eq:correlastic}) takes on negative values, rendering the present approximation unphysical at corresponding parameter values.\footnote{At the present level interaction corrections stem from the first term of the full Dyson series of interaction terms for non-local self energies. Summing the respective full series leads to an exclusively positive oscillatory result. Inclusion of these higher order terms requires a more sophisticated scheme to develop the reservoir region Green's function than provided by~(\ref{eq:altGtruncated}), which will be presented elsewhere. } The approach  is thus constrained by the condition\footnote{It has been demonstrated that, under these conditions, the relevant Green's function obtained from second order perturbation theory agrees with the exact Green's function from bosonization.\cite{Neuenhahn2008}}
\begin{align}
\label{eq:validity}
 \frac{\Delta x^2}{v^2} E^2(\omega) - \frac{\partial^2}{\partial \omega^2} E^2(\omega) \lesssim 1.
\end{align}
A plot of the corrected elastic current~(\ref{eq:correlastic}) is shown in Fig.~\ref{fig:ElasticCorrectedI} for the screened finite range interaction
\begin{align}
\label{eq:GaussianModel}
  \nu_q = \frac{\nu_0}{1+\lambda^2 q^2},
\end{align}
with constant $\nu_0$ and screening length $\lambda$. The poles of~(\ref{eq:GaussianModel}) are located at $q = \pm i / \lambda$ such that their contribution to the elastic correction is suppressed by $\exp \left( -\Delta x / \lambda \right)$, cf.~(\ref{eq:ElasticCorrContour}). Approximation~(\ref{eq:correlastic}) is thus valid as long as $\lambda \ll \Delta x$. Under this assumption, for~(\ref{eq:GaussianModel}), as well as for qualitatively similar screened interactions, condition~(\ref{eq:validity}) imposes restrictions on admissible values for interaction strength, dot separation, screening length, as well as injection energy. With~(\ref{eq:GaussianModel}), at~$\omega \approx v/\lambda$,~(\ref{eq:validity}) e.g.~turns into
 $\left(\nu_0/2 \pi v \right)^2 \left(\Delta x/ \lambda\right)^2 [\lambda (\omega - \mu_R) / v]^6/180 \lesssim 1$, where $\omega$ corresponds to either dot level $\omega_R$ or $\omega_L$. 
Corrections due to interactions cause the elastic current to diminish as the 
injection and detection energies increase
from the Fermi level.

\subsection{Inelastic contributions}
\label{sec:ContribInelastic}

Inelastic contributions to the current are determined by the third term of expansion~(\ref{eq:altGLangreth}) of the lesser and greater reservoir Green's function on the Keldysh contour. According to the Langreth rules, each of the electron lines in Fig.~\ref{fig:ExchPolGeneral} of the lesser self-energies in~(\ref{eq:altGLangreth}) corresponds to a lesser component Green's function, while the hole line corresponds to a greater component Green's function. For the greater self-energies, the opposite relations hold.  
Each of these lesser/greater tunneling Green's functions,
\begin{align}
\label{eq:GTkineticelaborate}
  \mathbf{G}^{T</>}_{\text{II}} = &\left(\mathbf{I} + \mathbf{G}^{Tr}_{\text{II}}  \mathbf{\Sigma}^r_\text{ITI} \right) \mathbf{g}^{</>}_{\text{II}} \left(\mathbf{I} + \mathbf{\Sigma}^a_\text{ITI} \mathbf{G}^{Ta}_{\text{II}}  \right) \nonumber \\ 
  &+ \mathbf{G}^{Tr}_{\text{II}} \mathbf{\Sigma}^{</>}_{\text{I}\text{D}_{\text{L}}\text{I}} \mathbf{G}^{Ta}_{\text{II}} + \mathbf{G}^{Tr}_{\text{II}} \mathbf{\Sigma}^{</>}_{\text{I}\text{D}_{\text{R}}\text{I}} \mathbf{G}^{Ta}_{\text{II}},
\end{align}
allows for equilibration in all of the setup's channels: while the first term in~(\ref{eq:GTkineticelaborate}) corresponds to equilibration in the reservoir itself (middle blue lines in Fig.~\ref{fig:ExchPolGeneral}), the second and third term are associated with equilibration in the source (green lines) and drain (red lines) leads, respectively. Thereby, both diagrams in Fig.~\ref{fig:ExchPolGeneral} are related to 27 distinct relaxation processes. 

\begin{figure}[b]
\subfloat[Direct electron current process\label{fig:1aProcess}]{%
    \includegraphics[width=.95\textwidth]{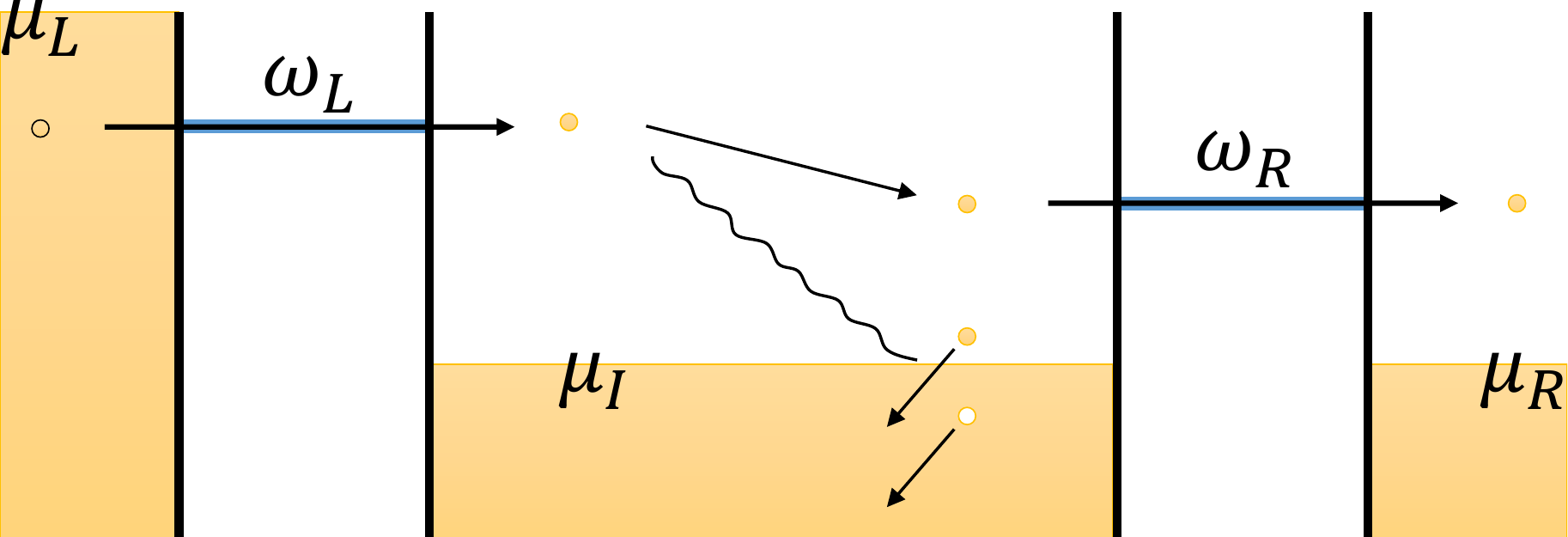}
  }
\vspace{1.5cm}
      \subfloat[Electron swap process\label{fig:1bProcess}]{%
    \includegraphics[width=.95\textwidth]{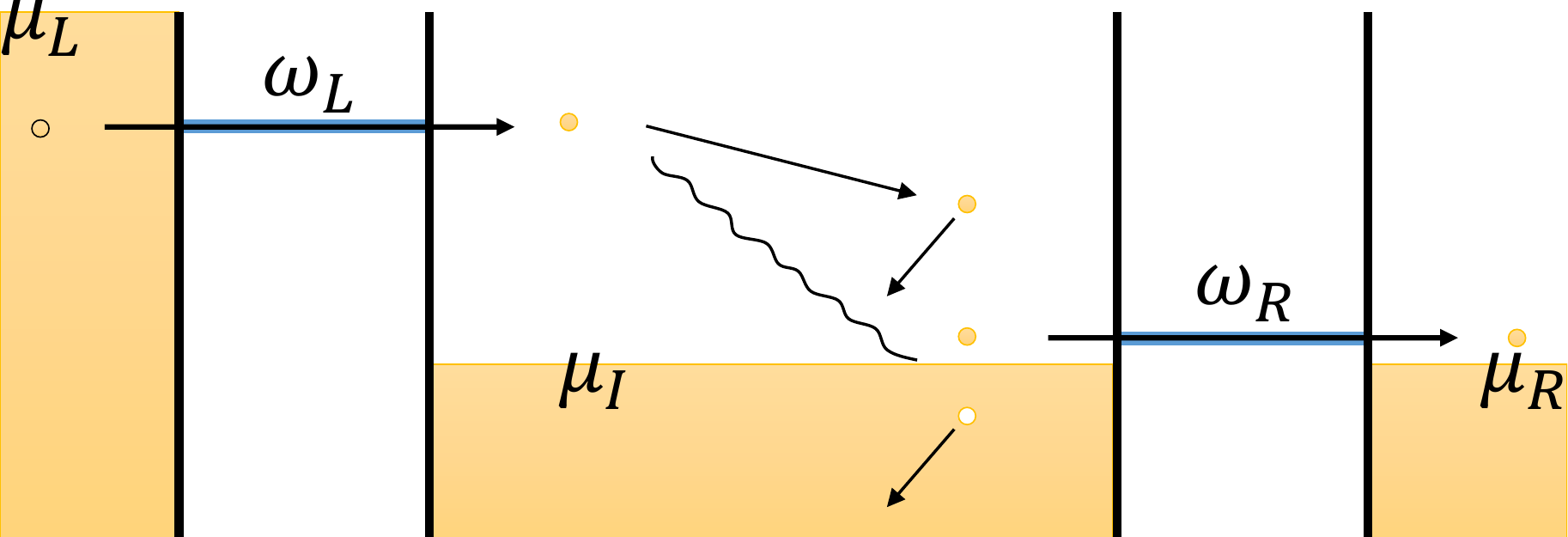}
  }
\vspace{1.5cm}
    \subfloat[Hole current process\label{fig:1cProcess}]{%
    \includegraphics[width=.95\textwidth]{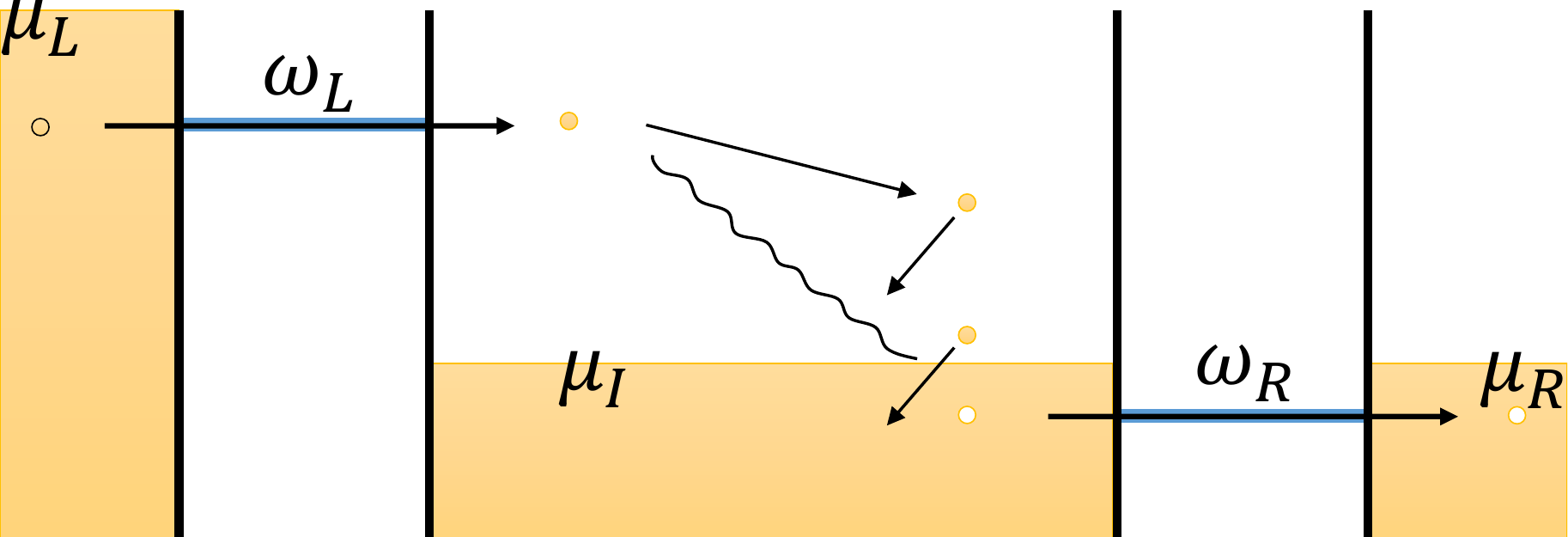}
  }
        \caption{Introducing interactions within the reservoir channel of the sample, three second order processes contribute to the inelastic current: (a) In the first process, an incoming electron generates an electron-hole pair, and subsequently enters the detector quantum dot at a lower energy. (b) In the second process, the electron of the pair enters the detector, while the initial electron equilibrates in the reservoir region. Due to indistinguishability of the electrons, processes (a) and (b) interfere. (c) In the third process, the hole of the pair can escape into the detector when the latter's energy is tuned below the Fermi level.}
  \label{fig:Processes}
\end{figure}

\begin{figure}[b]
    \includegraphics[width=.6\textwidth]{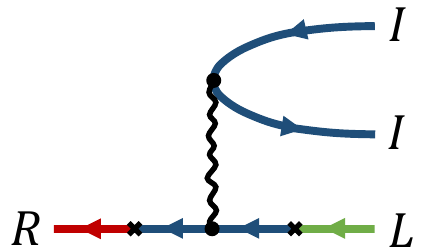}
    \caption{Diagram corresponding to the amplitude of the inelastic process depicted in Fig.~\ref{fig:1aProcess}. This process gives rise to contribution~(\ref{eq:1aexplicit}) to the current through the drain channel. }
  \label{fig:1a}
\end{figure}

These processes are revealed upon dividing the diagrams at the innermost Green's function lines, see e.g.~Fig.~\ref{fig:1a}. While the polarization diagrams are related to the probability of the individual processes (here the left half of each diagram corresponds to the complex conjugate of the respective right half), the exchange diagrams are related to interference terms due to indistinguishability of the electrons involved in these processes. 
 The diagrams are terminated by drain lines, since the current in this channel is calculated.

Diagrams are evaluated at zero temperature $T=0$, for equal chemical potentials in the reservoir and drain region $\mu_I = \mu_R$, and for delta function energy filter quantum dots, i.e.~upon invoking the approximation
\begin{align}
\label{eq:deltaT}
  \mathcal{T}_{L/R} (\omega) &= \frac{\Gamma^2}{(\omega-\omega_{L/R})^2 + \Gamma^2} \nonumber \\
                                                &\simeq  \pi \Gamma \delta(\omega - \omega_{L/R}).
\end{align}
Taking into account only one charge carrier transfer process per quantum dot,
as described in the last paragraph of Section~\ref{sec:interactions},
three of the 27 possible processes remain. In the remainder of the section, these processes will be evaluated.

\subsubsection{Direct electron current process, polarization diagram}
\label{sec:1a}

The first inelastic process considered is shown in the diagram in Fig.~\ref{fig:1a}.  
The dynamics which corresponds to this diagram is depicted in Fig.~\ref{fig:1aProcess}: an electron from the source lead enters through the emitter quantum dot into the reservoir region. In the reservoir region, this electron dissipates energy upon creation of an electron-hole pair. While the initial electron is transmitted through the detector dot into the drain lead, the electron and hole of the pair equilibrate in the reservoir region. 

The current generated by the diagram expressed in terms of Keldysh Green's functions~(\ref{eq:I1a}) is contained in Appendix~\ref{app:InelasticContributions}.
 Collecting tunneling amplitudes which combine to the product of the transition probabilities 
 $\mathcal{T}_{L}$ and $\mathcal{T}_{R}$ 
 through the left and right dot at energies $\omega + \omega''$ and $\omega$, respectively, this contribution becomes
\begin{align}
\label{eq:1afs}
I_R^{\text{1a}} = &-\frac{e}{(2\pi)^3} \int_{-\infty}^{\infty}\!\!\!d\omega'' \int_{-\infty}^{\infty}\!\!\!d\omega' \int_{-\infty}^{\infty}\!\!\!d\omega \mathcal{T}_L(\omega + \omega'') \mathcal{T}_R(\omega) \nonumber\\
 \times&\big[ f_R(\omega) \left(f_L(\omega + \omega'') -1 \right) f_I\left( \omega' + \omega'' \right) \left( f_I\left( \omega' \right) - 1 \right) \nonumber \\
& \qquad \qquad \qquad- \text{``}f \leftrightarrow (f-1)\text{''} \big] |\Upsilon(\omega'')|^2, 
\end{align}

\noindent where
\begin{align}
\label{eq:upsilon}
  \Upsilon(\omega'') &= -\frac{\nu_{\omega''/v}}{2\pi} \int_{-\infty}^{\infty} dk \frac{\exp\left( i \frac{\Delta x}{v} \left(vk- \omega \right) \right)}{\left( vk-\omega -i\delta \right)^2 } \nonumber \\
                     &=\frac{\nu_{\omega''/v}}{v}\frac{\Delta x}{v}.
\end{align}
For delta-like filters~(\ref{eq:deltaT}) and $\mu_I= \mu_R$, inelastic contribution~(\ref{eq:1afs}) turns into
\begin{align}
\label{eq:1aexplicitT}
&I_R^{\text{1a}} = \frac{e}{2\pi} \frac{\Gamma^2}{4} \left(\omega_L-\omega_R \right) |\Upsilon(\omega_L-\omega_R)|^2 \nonumber \\ 
                                   &\times \frac{1}{4} \text{csch}\left[ \frac{\beta}{2}  (\omega_L-\omega_R) \right] \text{sech}\left[ \frac{\beta}{2} (\mu_L-\omega_L) \right] \nonumber \\
                                   &\quad \times  \text{sech}\left[ \frac{\beta}{2} (\mu_R-\omega_R) \right] \sinh\left[ \frac{\beta}{2} (\mu_L-\mu_R) \right].
\end{align}
For infinite temperature,~(\ref{eq:1aexplicitT}) becomes 
\begin{align}
\label{eq:1aexplicitInf}
\!\!\!\!\!\!\!\!\!\!\!\!\!\!\!\!\!\!\!\!\!\!\!\!\!\!\!\!\!\!\!\!\!\!\!\!\!\!\!\!\!\!\!\!\!\!\!\!\!\! I_R^{\text{1a}} = \frac{e}{2\pi} \frac{\Gamma^2}{4} \frac{\mu_L-\mu_R}{4} |\Upsilon(\omega_L-\omega_R)|^2,
\end{align}
whereas at zero temperature we find
\begin{align}
\label{eq:1aexplicit}
&I_R^{\text{1a}} = \frac{e}{2\pi} \frac{\Gamma^2}{4} \left(\omega_L-\omega_R \right) |\Upsilon(\omega_L-\omega_R)|^2 \nonumber \\ 
                                   &\!\!\times \left[ \theta(\mu_R- \omega_R) - \theta(\mu_L- \omega_R) \right] \left[ \theta(\omega_R- \omega_L) - \theta(\mu_L- \omega_L) \right].
\end{align}
The first line of~(\ref{eq:1aexplicit}) contains spatial and energetic dependencies of the current. 
The second line of~(\ref{eq:1aexplicit}) describes the outline of triangle \textbf{\rom{1}} in Fig.~\ref{fig:current}, in the Emitter-Detector energy space. 

\begin{figure}[t]
    \includegraphics[width=.95\textwidth]{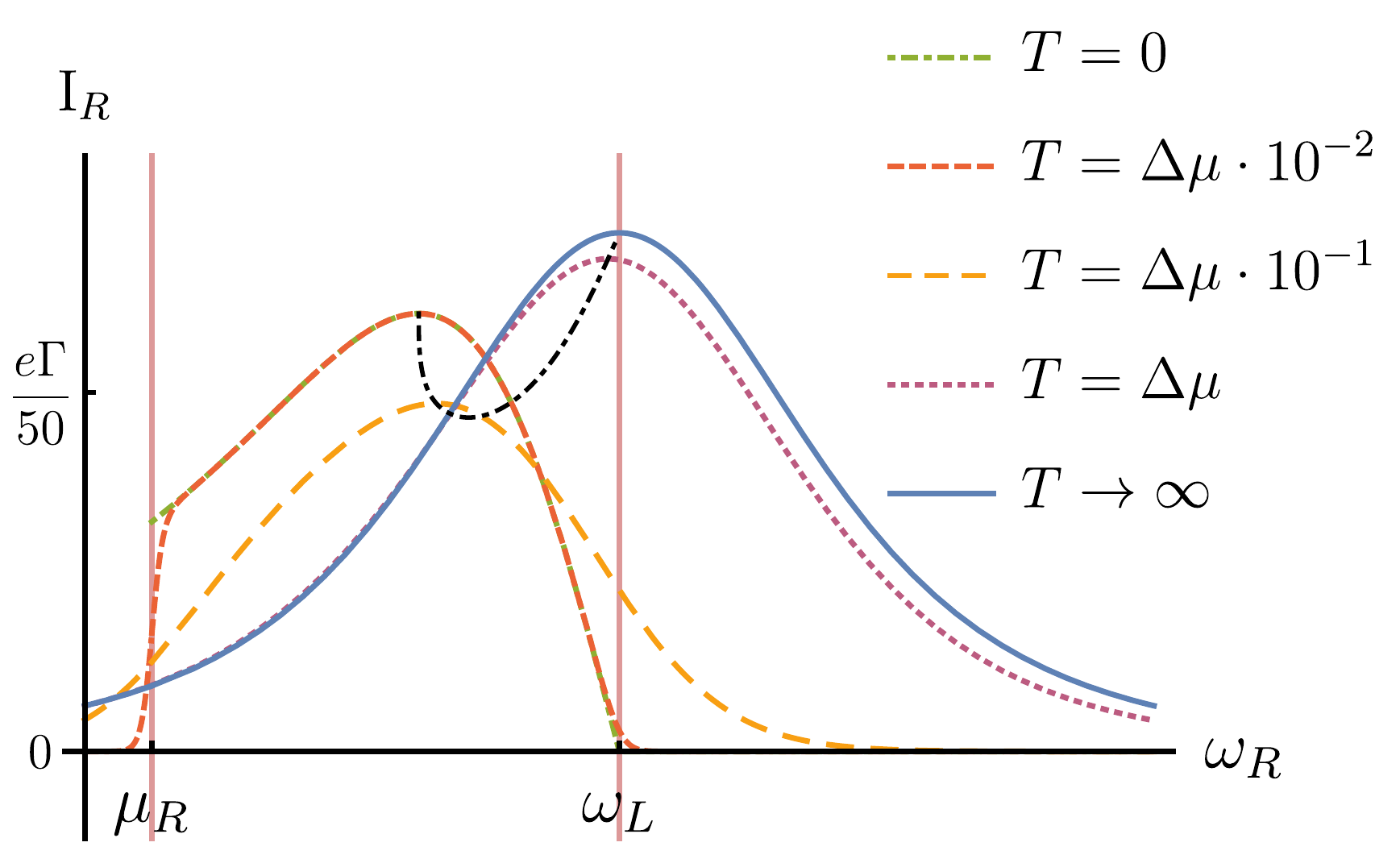}
    \caption{Temperature dependence of inelastic contribution~(\ref{eq:1aexplicitT}), stemming from the process depicted in Fig.~\ref{fig:1aProcess}, at fixed emitter energy. At temperatures comparable to the experimental value (red dashed), the current is in good agreement with its value at $T=0$ (green dash-dotted), except within a range of $T$ to triangle outline~(\ref{eq:1aexplicit}). The black dash-dotted line indicates the maximum of the current as T is varied between $0$ and $\infty$. (Parameters are $\mu_L = \mu_R + 400\Gamma$, $\omega_L = \mu_R + 350\Gamma$, $v = 260\lambda\Gamma$, $\nu_0 = 720\lambda\Gamma$, $x =8\lambda$.)   }
  \label{fig:1aTemp}
\end{figure}
Figure~\ref{fig:1aTemp} shows the current~(\ref{eq:1aexplicitT}) at given bias $\Delta \mu = \mu_L-\mu_R$ and emitter energy $\omega_L$ for several values of the temperature, for the model interaction~(\ref{eq:GaussianModel}). At temperatures comparable to the experimental value, $T = \Delta \mu \cdot 10^{-2}$ (red line), except within a range of $T$ to the outline of triangle \textbf{\rom{1}}, the current is in good agreement with its value for zero temperature (green line). All following diagrams are thus evaluated at $T=0$. At infinite temperature (blue line), the current is proportional to the bias $\Delta \mu$, and its emitter-detector-energy dependence is completely determined by the form of the interaction, cf.~(\ref{eq:upsilon}) and~(\ref{eq:1aexplicitInf}).

\subsubsection{Electron swap process, polarization diagram}
\label{sec:1b}

\begin{figure}[t]
    \includegraphics[width=.6\textwidth]{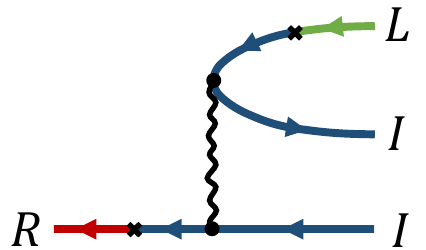}
    \caption{Diagram corresponding to the process depicted in Fig.~\ref{fig:1bProcess}, in which the source electron is swapped for an electron from the Fermi sea, subsequently entering the detector quantum dot.}
  \label{fig:1b}
\end{figure}

The contribution of the second inelastic process is determined by the diagram displayed in Fig.~\ref{fig:1b}. In contrast to the previous physical process considered in Fig.~\ref{fig:1aProcess},  here the electron of the electron-hole pair generated in the reservoir region enters the detector quantum dot, as depicted in Fig.~\ref{fig:1bProcess}.
Since the involved electrons are indistinguishable, the two processes interfere. The information about this interference is contained in the diagrams of the exchange self-energy, as will become apparent in section~\ref{sec:2a2b}.

Analogously to section~\ref{sec:1a}, the contribution of the diagram to the current, expressed in terms of Keldysh Green's functions~(\ref{eq:I1b}), turns into 
\begin{align}
\label{eq:1bexplicit}
&I_R^{\text{1b}} = \frac{e}{2\pi} \frac{\Gamma^2}{4} \int_{\mu_I-\omega_L}^{\mu_I-\omega_R}\!\!\!\!\!\! d\omega'' \, |\Xi(\omega'')|^2 \nonumber \\
                                   &\!\!\times \left[ \theta(\mu_R- \omega_L) - \theta(\mu_L- \omega_L) \right] \left[ \theta(\mu_R- \omega_R) - \theta(\omega_L- \omega_R) \right],
\end{align}
where
\begin{align}
\label{eq:Xi}
 \Xi(\omega'') =  -\frac{1}{2\pi} \int_{-\infty}^{\infty} dq \, \nu_q \frac{\exp\left( i \frac{\Delta x}{v} \left(vq- \omega'' \right) \right)}{\left( vq-\omega'' -i\delta \right)^2}.
\end{align}
The second line in~(\ref{eq:1bexplicit}) once again marks the outline of triangle \textbf{\rom{1}} in Fig.~\ref{fig:current}.\footnote{The combinations of $\theta$ functions in~(\ref{eq:1aexplicit}) and~(\ref{eq:1bexplicit})  can be transformed into one another upon repeated application of $\theta(x-a)\theta(b-x) = \theta(b-a)[\theta(b-x)-\theta(a-x)] $}

\subsubsection{Hole current process, polarization diagram}
\label{sec:1c}

\begin{figure}[t]
    \includegraphics[width=.65\textwidth]{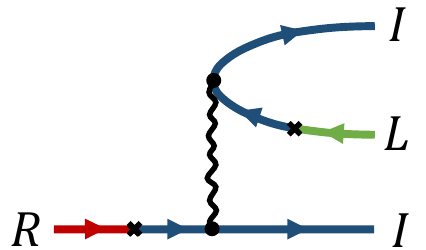}
    \caption{Diagram corresponding to the process in Fig.~\ref{fig:1cProcess}, in which the hole of the electron-hole pair that is generated by the source electron escapes into the detector quantum dot.}
  \label{fig:1c}
\end{figure}

The third inelastic contribution is described by the diagram displayed in Fig.~\ref{fig:1c}. The corresponding physical process is depicted in Fig.~\ref{fig:1cProcess}. In this process, the hole of the electron-hole pair, generated by the source electron, can tunnel into the detector quantum dot when the latter's resonant level is tuned below the Fermi sea. Thereby, a current of the opposite sign compared to the previously considered inelastic contributions is generated.

Similarly to the processes in sections~\ref{sec:1a} and~\ref{sec:1b}, the Keldysh Green's function expression~(\ref{eq:I1c}) simplifies to 
\begin{align}
\label{eq:1cexplicit}
I_R^{\text{1c}} = -&\frac{e}{2\pi} \frac{\Gamma^2}{4} \int_{\mu_I-\omega_R}^{\omega_L-\mu_I}\!\!\!\!\!\! d\omega'' \, |\Xi(\omega'')|^2 \nonumber \\
                                   &\times \left[ \theta(\mu_R- \omega_R) - \theta(2\mu_I - \mu_L- \omega_R) \right] \nonumber \\
                                   &\qquad\times  \left[ \theta(\mu_L- \omega_L) - \theta(2\mu_I-\omega_R- \omega_L) \right].
\end{align}

The second and third line of~(\ref{eq:1cexplicit}) in this case confine the current to triangle \textbf{\rom{2}} in Fig.~\ref{fig:current}. As a function of the detector energy $\omega_R$,  contribution~(\ref{eq:1cexplicit}) constitutes the point reflection of~(\ref{eq:1bexplicit}) at the chemical potential $\mu_R$.  

\begin{figure*}[t]
\subfloat[$I_{R}^{\text{2a}}$\label{fig:2a}]{%
    \includegraphics[width=.46\textwidth]{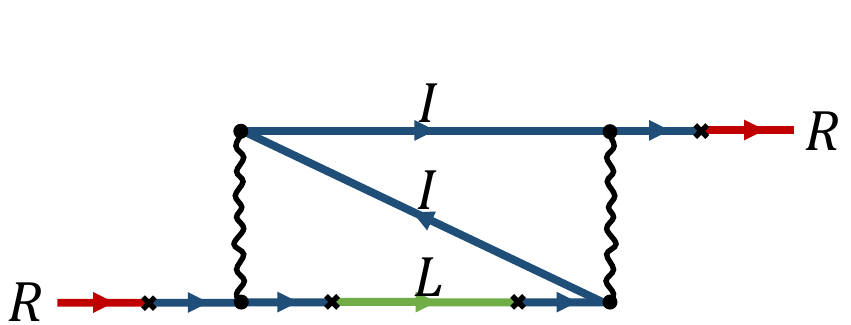}
  }
\hspace{.9cm}
      \subfloat[$I_{R}^{\text{2b}}$\label{fig:2b}]{%
    \includegraphics[width=.46\textwidth]{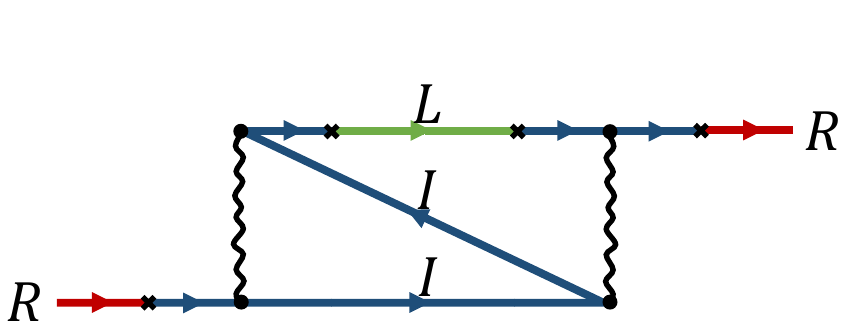}
  }
        \caption{The exchange diagrams for the electron current are obtained upon combining the process diagrams in Figs.~\ref{fig:1a} and~\ref{fig:1b}. The diagrams describe interference of these processes. }
  \label{fig:2a2b}
\end{figure*}

\subsubsection{Direct and swap electron processes, exchange diagram}
\label{sec:2a2b}

The exchange diagrams of direct and swap electron processes are shown in Fig.~\ref{fig:2a2b}. The diagrams can be obtained by joining the diagram halves in Figs.~\ref{fig:1a} and \ref{fig:1b} by matching respective Green's function lines, which indicates that the exchange diagrams correspond to interference of the corresponding processes, as is confirmed by evaluation of the expressions below. 

\begin{figure}[b]
    \includegraphics[width=\textwidth]{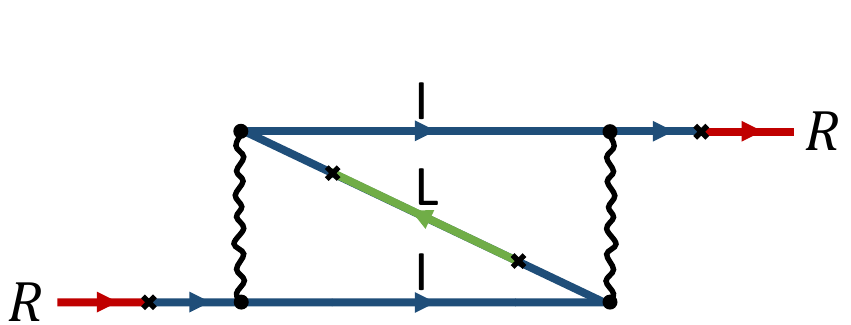}
    \caption{Exchange diagram corresponding to the hole current process depicted in Fig.~\ref{fig:1cProcess}. The diagram reduces the current as a consequence of the indistinguishability of electrons.}
  \label{fig:2c}
\end{figure}

The diagrams constitute the complex conjugate of one-another, so it is sufficient to determine the contribution of the diagram in Fig.~\ref{fig:2a}.
After employing the same approximations as in the previous diagrams to the Keldysh Green's functions expression~(\ref{eq:I2a}), the contribution simplifies to
\begin{align}
\label{eq:2aexplicit}
&I_R^{\text{2a}} = -\frac{e}{2\pi} \frac{\Gamma^2}{4} \Upsilon(\omega_L-\omega_R) \int_{\mu_I-\omega_L}^{\mu_I-\omega_R}\!\!\!\!\!\! d\omega'' \, \Xi^*(\omega'') \nonumber \\
                                   &\!\!\times  \left[ \theta(\mu_R- \omega_R) - \theta(\mu_L- \omega_R) \right] \left[ \theta(\omega_R- \omega_L) - \theta(\mu_L- \omega_L) \right].
\end{align}
For completeness, we also state the corresponding expression for the diagram in Fig.~\ref{fig:2b},
\begin{align}
\label{eq:2bexplicit}
&I_R^{\text{2b}} = -\frac{e}{2\pi} \frac{\Gamma^2}{4} \Upsilon^*(\omega_L-\omega_R) \int_{\mu_I-\omega_L}^{\mu_I-\omega_R}\!\!\!\!\!\! d\omega'' \, \Xi(\omega'') \nonumber \\
                                   &\!\!\times  \left[ \theta(\mu_R- \omega_R) - \theta(\mu_L- \omega_R) \right] \left[ \theta(\omega_R- \omega_L) - \theta(\mu_L- \omega_L) \right].
\end{align}
From the explicit expressions at given energies, it is apparent that (\ref{eq:2aexplicit}) and~(\ref{eq:2bexplicit}) correspond to the interference terms of the processes generating~(\ref{eq:1aexplicit}) and~(\ref{eq:1bexplicit}). The latter processes are obtained from one another by the exchange of an electron, such that their interference terms contribute with a negative sign.

\subsubsection{Hole current process, exchange contribution}

The last remaining diagram constitutes the exchange contribution of the hole current process of section~\ref{sec:1c}. The diagram is displayed in Fig.~\ref{fig:2c} and generates the Keldysh space expression~(\ref{eq:I2c}), which subsequently simplifies to
\begin{align}
\label{eq:2cexplicit}
I_R^{\text{2c}} = &\frac{e}{2\pi} \frac{\Gamma^2}{4} \int_{\mu_I-\omega_R}^{\omega_L-\mu_I}\!\!\!\!\!\! d\omega'' \, \Xi^*(\omega'')\Xi(\omega_L-\omega_R-\omega'') \nonumber \\
                                   &\times \left[ \theta(\mu_R- \omega_R) - \theta(2\mu_I - \mu_L- \omega_R) \right] \nonumber \\
                                   &\qquad\times  \left[ \theta(\mu_L- \omega_L) - \theta(2\mu_I-\omega_R- \omega_L) \right],
\end{align}
 contributing with the opposite sign of~(\ref{eq:1cexplicit}).

For contact interaction, $\nu_q \equiv \text{const}$, we find $\Xi \equiv \Upsilon$, such that all diagrams cancel. This reflects that two electrons cannot occupy the same position in the reservoir channel, such that no relaxation by means of contact interaction is possible. Intra-channel relaxation thus requires  finite range interactions between the charge carriers.

\begin{figure}[b]
    \includegraphics[width=.95\textwidth]{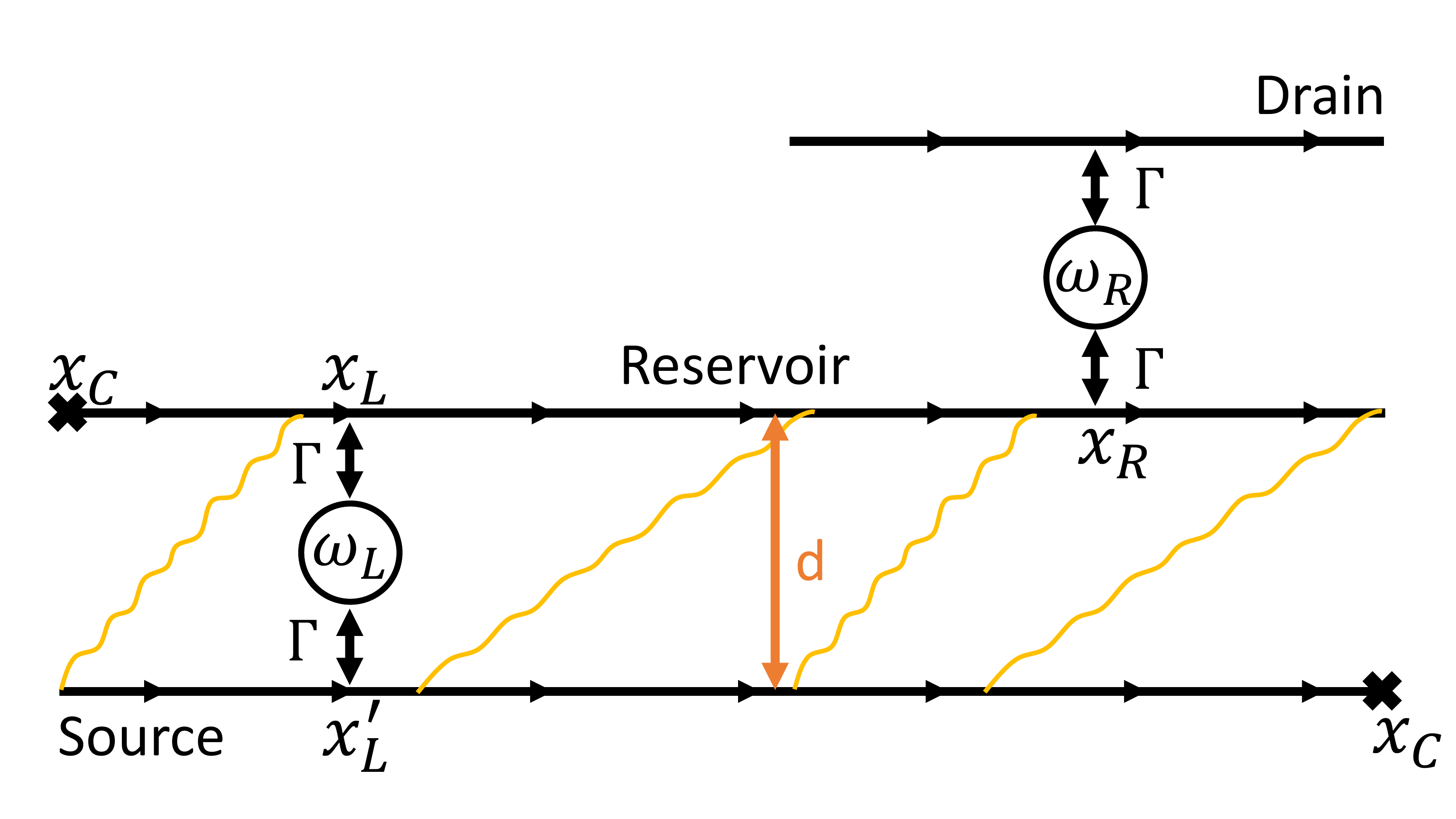}
    \caption{Simple model for interchannel interaction between reservoir and source. To close the channels into a single edge with given chirality, a point $x_C$ downstream of the emitter at $x_L'$ in the source channel is identified with a point upstream of the emitter at $x_L$ in the reservoir channel. The interaction is subsequently defined to be maximal when the distance of electrons in the reservoir channel measured from $x_L$ coincides with the distance of electrons in the source channel measured from $x_L'$.}
  \label{fig:InterchannelModel}
\end{figure}

\section{Interactions between reservoir region and source lead}
\label{sec:HVIL}

Taking into account tunneling and interactions between several chiral channels, a consistent treatment requires to artificially close the system into a single edge with fixed chirality.~\cite{PhysRevB.65.153304} 
Upon inspection of the micrograph of the experimental sample shown in Fig.~\ref{fig:sample}, one finds that the one-dimensional channels of the source and reservoir region partly copropagate and partly counterpropagate. In a simplified model, we treat the source and reservoir components of the sample as parallel channels with the same chirality, see Fig.~\ref{fig:InterchannelModel}. To achieve the most basic consistent treatment, the channels are connected by identifying a point in the source channel downstream of the emitter quantum dot with a point in the reservoir channel upstream of the emitter, denoted by $x_C$. Interaction of the charge carriers on the channels is further required to be maximal when their horizontal coordinates in Fig.~\ref{fig:InterchannelModel} coincide, where coordinates in the reservoir and source channels are measured from the dot coordinates $x_L$ and $x_L'$, respectively.

Following these considerations, the interchannel interaction Hamiltonian takes the form\footnote{Interactions between the reservoir region and the drain channel are not taken into account, since at equal chemical potentials $\mu_I=\mu_R$ and $T=0$ no inelastic processes are generated.}
\begin{align}
H_V^{IL} &= \frac{1}{2} \int_{x_C}^{\infty} dx_{\text{Res}} \int_{-\infty}^{x_C} dx_{\text{S}} \nu^d (x_{\text{Res}} - x_L -  x_{\text{S}} + x_L') \nonumber \\
                &\qquad \qquad \qquad  \times \hat{\psi}_{x_{\text{Res}}}^\dagger \hat{\psi}_{x_{\text{S}}}^\dagger \hat{\psi}_{x_{\text{S}}}^{\vphantom{\dagger}}\hat{\psi}_{x_{\text{Res}}}^{\vphantom{\dagger}}
\end{align}
as expressed in terms of spatial coordinates. The superscript of the interaction matrix element in position space $\nu^d$ indicates accounting for the additional vertical channel separation $d$ in Fig.~\ref{fig:InterchannelModel}. Taking $x_C$ to infinity in the source channel and to negative infinity in the reservoir channel, reflecting that the point in which the channels close is located far away from the emitter quantum dot, the interaction Hamiltonian turns into
\begin{align}
\label{eq:HVIL}
  H_V^{IL} = \frac{1}{2\Omega} \sum_{k,k_L,q} \nu_q^{d\vphantom{\dagger}} \text{e}^{i q (x_L-x_L')} \hat{r}^\dagger_{k-q} \hat{l}^\dagger_{k_L+q} \hat{l}_{k_L}^{\vphantom{\dagger}} \hat{r}_{k}^{\vphantom{\dagger}}.
\end{align}

\begin{figure}[t]
    \includegraphics[width=.97\textwidth]{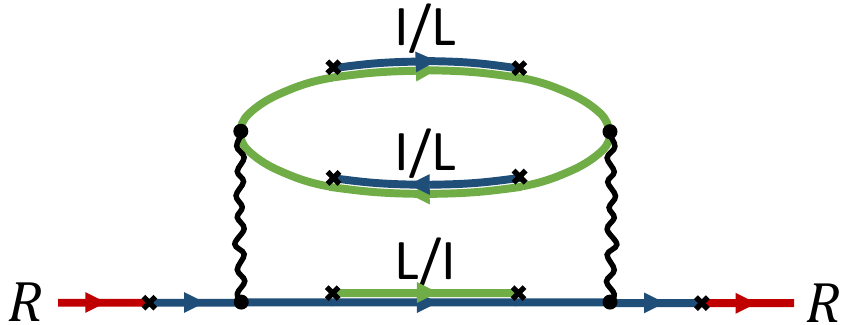}
    \caption{Polarization contribution generated by interactions between charge carriers in the source lead and the reservoir region, showing relevant tunneling processes only.}
  \label{fig:PolGenIL}
\end{figure}
The development of the Dyson equation for inter-channel interaction~(\ref{eq:HVIL}) and the assignment of the respective Green's function components in Keldysh space proceeds similarly as in Section~\ref{sec:GFInteracting}. In contrast to that section, inter-channel interactions no longer allow for the same concise representation of all possible processes as is provided by the truncated Dyson equation~(\ref{eq:altGtruncated}). Instead, the diagrams of all of these possible tunneling processes have to be considered at the second order of interactions, to be developed in Keldysh space. Afterwards, similar approximations as in the case of intra-channel interactions are invoked. 

In the diagrams generated by~(\ref{eq:HVIL}), the phase proportional to $x_L-x_L'$ shifts the tunneling coordinate $x_L'$ in the source channel to the coordinate $x_L$  in the reservoir channel, such that all currents become functions of the distance $\Delta x = x_L-x_R$, as it is the case for intrachannel interaction~(\ref{eq:HV}), compare~(\ref{eq:upsilon}) and~(\ref{eq:Xi}).

The interaction~(\ref{eq:HVIL}) generates the second order diagram shown in Fig.~\ref{fig:PolGenIL}, in which only relevant tunneling processes are displayed. Polarization and exchange diagrams featuring both inter- and intrachannel interaction are energetically strongly suppressed. 

In the polarization diagram, the Green's function lines in the closed Fermion loop here correspond to tunneling Green's functions of the source lead, which develop with the bare Hamiltonian $H_0$ as well as with the tunneling Hamiltonian $H_T$.

\subsection{Corrections to elastic current}

The interaction between source and reservoir channel~(\ref{eq:HVIL}) contributes a further polarization-diagram term to the interacting self-energy~(\ref{eq:sigma}), compare~(\ref{eq:Epol}),  and thus to the corrected elastic current~(\ref{eq:correlastic}). This term is given by
\begin{align}
\label{eq:EpolIL}
   E_{\text{pol},\text{IL}}^{2}(\omega) = \frac{1}{(2\pi v)^2} \int_{0}^{\mu_I -\omega} \! \!\!\!\!\!\!\! d\omega'\, \omega'  \left({\nu_{\omega'/v}^{d}}\right)^2.
\end{align}
Apart from the additional distance  across the source quantum dot accounted for by $\nu_q^{d\vphantom{\dagger}}$, the structure of~(\ref{eq:Epol}) and~(\ref{eq:EpolIL}) is identical. The latter term thus adds to the decay of the inelastic peak as the distance between the dots or the distance of the quantum dots' resonant levels from the Fermi level is increased. Exchange diagrams do not contribute for the relevant tunneling processes.
\footnote{The interaction~(\ref{eq:HVIL}) also introduces a correction of the local (i.e.~independent of tunneling phases) transition amplitude
$
\label{eq:altGLLLL}
   \mathbf{t}_{ \text{D}_\text{L} \text{L}} \mathbf{g}^{V<}_{\text{LL}} \mathbf{t}_{\text{L}\text{D}_\text{L} }
   \simeq \mathbf{t}_{\text{D}_\text{L} \text{L}} \Big[\mathbf{g}^{<}_{\text{LL}} + \mathbf{g}^{<}_{\text{LL}} \mathbf{\Sigma}^{Va}_{\text{LVL}} \mathbf{g}^{a}_{\text{LL}}      
  + \mathbf{g}^{r}_{\text{LL}} \mathbf{\Sigma}^{V<}_{\text{LVL}} \mathbf{g}^{a}_{\text{LL}} + \mathbf{g}^{r}_{\text{LL}} \mathbf{\Sigma}^{Vr}_{\text{LVL}} \mathbf{g}^{<}_{\text{LL}} \Big] \mathbf{t}_{\text{L} \text{D}_\text{L}}
$
within the source lead, in which $ \mathbf{g}^{V}_{\text{LL}}$ denotes the Green's function including interactions but not tunneling. In the perturbative approach, the last three terms of the above amplitude diverge. Since the transition amplitude is local, a full solution only changes  occupation probability and density of states in the left lead, without immediate influence on transfer processes and relaxation between the quantum dots. Such local corrections are therefore  neglected.}

\begin{figure}[t]
    \includegraphics[width=.6\textwidth]{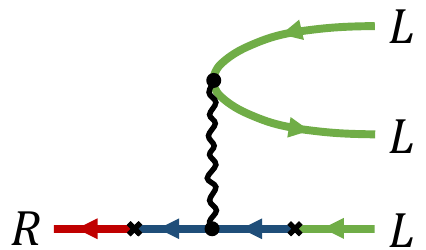}
    \caption{Diagram describing the process in which the source electron, after tunneling into the reservoir region, creates an electron-hole pair in the source lead, before entering the detector quantum dot. The diagram is structurally equivalent to the diagram in Fig.~\ref{fig:1a}, and reflects that the source lead provides a further decay channel when interactions between source and reservoir are taken into account.}
  \label{fig:3a}
\end{figure}

\subsection{Inelastic contributions}

Contributions to the inelastic current due to the interaction between source and reservoir channel~(\ref{eq:HVIL}) are described by the polarization diagram in Fig.~\ref{fig:PolGenIL}. To evaluate the diagram, the tunneling Green's functions of the source lead which appear in the closed Fermion loop have to be expanded on the Keldysh contour. This expansion is carried out in Appendix~\ref{app:sourceGF}. Here equilibration in the self-energy Green's function lines  for the relevant tunneling processes is restricted to the source and reservoir region, as indicated in Fig.~\ref{fig:PolGenIL}, compare Fig.~\ref{fig:PolGen}.

 After collection of terms which combine to the transition probability~(\ref{eq:transition}) through the left and right quantum dot, the contributions generated by the interaction~(\ref{eq:HVIL}) are identical in structure to the polarization diagram processes of sections~\ref{sec:1a},~\ref{sec:1b}, and~\ref{sec:1c}, respectively, with the exception of two properties: the additional distance $d$ across the emitter quantum dot, see Fig.~\ref{fig:InterchannelModel}, is accounted for by the replacement of the momentum space interaction matrix element $\nu_q$ by $\nu_q^{d}$, and tunneling through the emitter quantum dot on the Fermion loop Green's function lines in Fig.~\ref{fig:PolGenIL} here fixes the energy of the tunneling electron in the reservoir region instead of in the source lead. In the following we find that the latter distinction leads to the generation of triangles \textbf{\rom{3}} and \textbf{\rom{4}} observed in the experimental data displayed in Fig.~\ref{fig:current}. 

 \begin{figure}[t]
    \includegraphics[width=.6\textwidth]{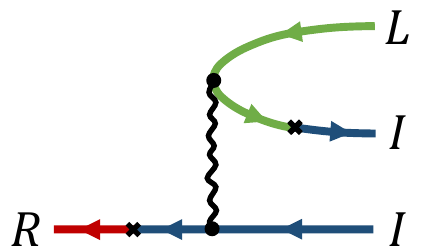}
    \caption{Diagram corresponding to the process in Fig.~\ref{fig:3bProcess}, in which the source electron interacts with the reservoir region where an electron-hole pair is created. The electron of the pair subsequently passes through the detector. The diagram creates current in  triangle \textbf{\rom{3}} in the experimental data shown in Fig.~\ref{fig:current}.}
  \label{fig:3b}
\end{figure}

\begin{figure}[b]
      \subfloat[Inverted triangle electron swap process\label{fig:3bProcess}]{%
    \includegraphics[width=.95\textwidth]{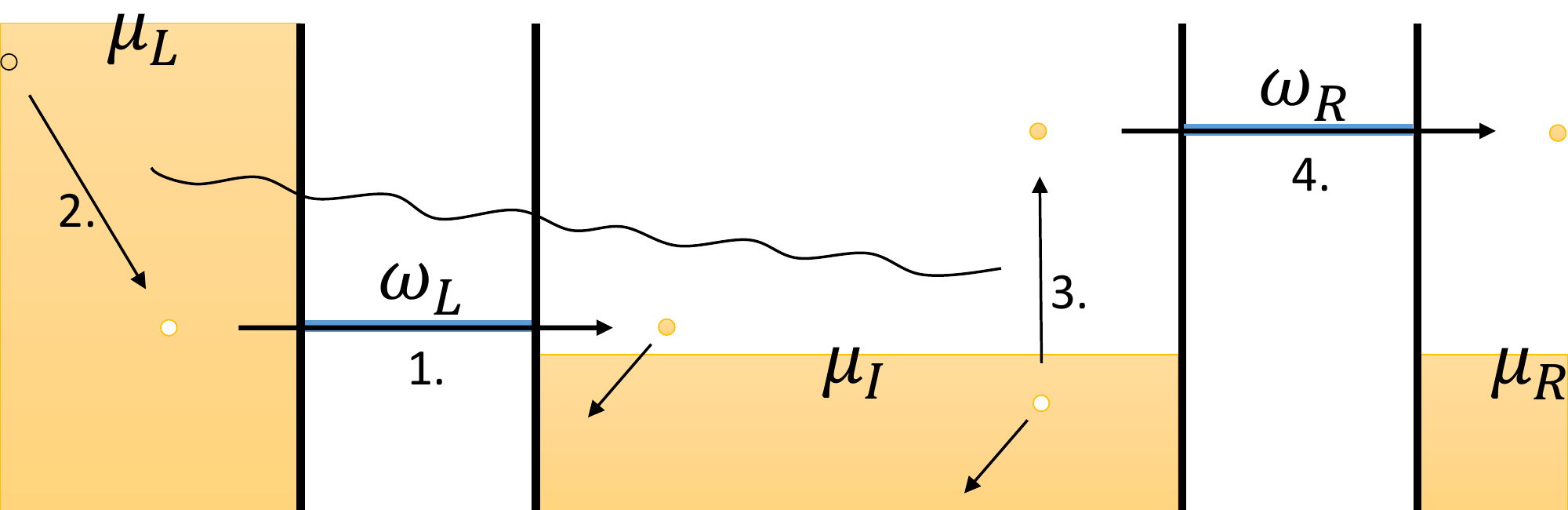}
  }
\vspace{1.5cm}
    \subfloat[Inverted triangle hole current process\label{fig:3cProcess}]{%
    \includegraphics[width=.95\textwidth]{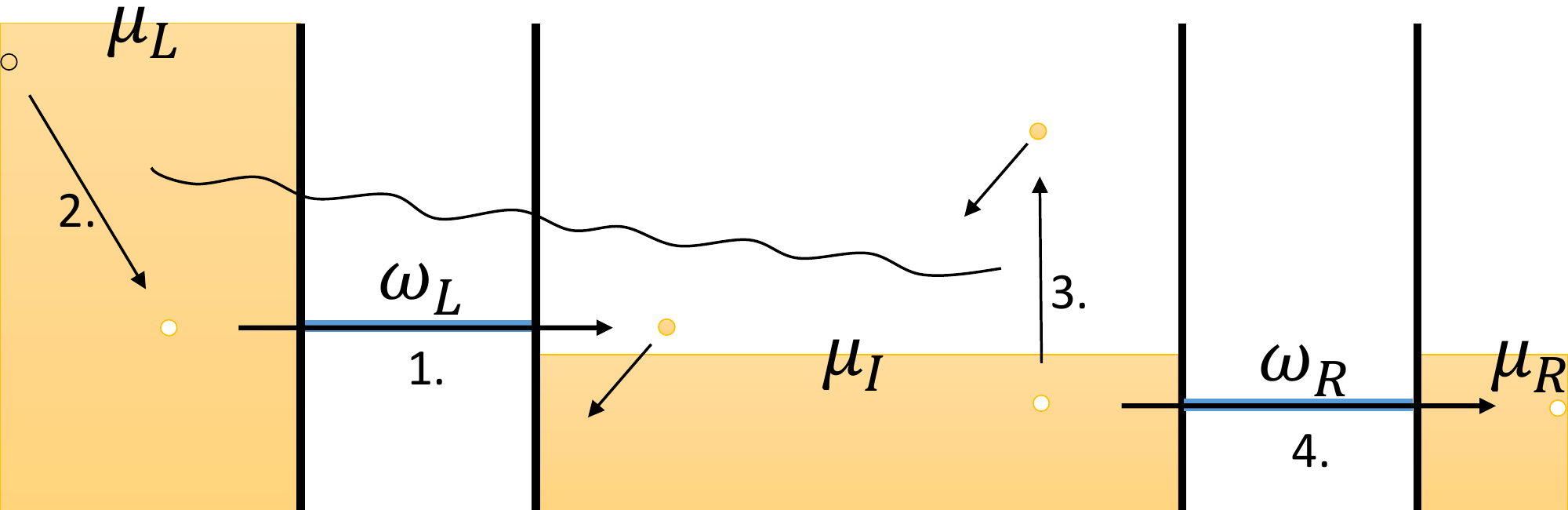}
  }
        \caption{Introducing interactions between reservoir and source channels of the sample, three further inelastic second order processes generate inelastic current: the first process (not shown) is equivalent to the process displayed in Fig.~\ref{fig:1aProcess}, with the exception that the electron-hole pair is generated in the source lead. (a) In the second process, \textbf{1.}~an electron of the source lead enters the reservoir region. This electron is  \textbf{2.}~subsequently replaced by a further source electron dissipating energy. This energy \textbf{3.}~creates an electron-hole pair in the reservoir region. The electron of the pair then \textbf{4.}~enters the drain lead, where the absolute distance of the detector quantum dot energy from the Fermi level can be larger than the corresponding distance for the emitter. This process generates the upper left triangle in Fig.~\ref{fig:current}. (b) In the third process, the hole of the pair enters the drain when the detector energy is located below the Fermi level.}
  \label{fig:Processes}
\end{figure}

\begin{figure}[t]
    \includegraphics[width=.6\textwidth]{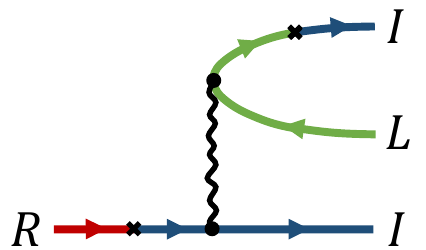}
    \caption{Diagram corresponding to the process of Fig.~\ref{fig:3cProcess}. In this process, the hole of an electron-hole pair, created in the reservoir by a virtual photon from the source, enters the detector. The diagram generates current in triangle \textbf{\rom{4}} in Fig.~\ref{fig:current}.}
  \label{fig:3c}
\end{figure}

\subsubsection{Direct electron current process}

Taking into account interactions between reservoir and source~(\ref{eq:HVIL}), the first diagram under consideration is shown in Fig.~\ref{fig:3a}. 
The expression generated by the diagram is identical to the contribution of the diagram in Fig.~\ref{fig:1a}, with the exception of the modification due to the additional distance across the emitter quantum dot. The present diagram therefore contributes
\begin{align}
\label{eq:3aexplicit}
&I_R^{\text{3a}} = \frac{e}{2\pi} \frac{\Gamma^2}{4} \left(\omega_L-\omega_R \right) |\Upsilon_d(\omega_L-\omega_R)|^2 \nonumber \\
                                   &\!\!\times \left[ \theta(\mu_R- \omega_R) - \theta(\mu_L- \omega_R) \right] \left[ \theta(\omega_R- \omega_L) - \theta(\mu_L- \omega_L) \right],
\end{align}
where
\begin{align}
\label{eq:upsilonresult}
  \Upsilon_d(\omega'') =\frac{\nu^d_{\omega''/v}}{v}\frac{\Delta x}{v},
\end{align}
compare~(\ref{eq:upsilon}) and~(\ref{eq:1aexplicit}). The diagram in Fig.~\ref{fig:3a} thus adds a further term to the current generated in triangle \textbf{\rom{1}} in Fig.~\ref{fig:current}, since the interaction with the source lead provides an additional decay channel for reservoir electrons.

\subsubsection{Inverted triangle electron swap}

The second diagram generated by~(\ref{eq:HVIL}) is shown in Fig.~\ref{fig:3b}. The physical process  corresponding to the diagram  is displayed in Fig.~\ref{fig:3bProcess}. Here, a virtual photon from the source generates an electron-hole pair in the reservoir region. The electron of the pair subsequently enters the drain channel via the detector quantum dot. The diagram is structurally equivalent to the diagram in Fig.~\ref{fig:1c} and contributes
\begin{align}
I_R^{\text{3b}} &= -\frac{e}{(2\pi)^3} \int_{-\infty}^{\infty}\!\!\!d\omega'' \int_{-\infty}^{\infty}\!\!\!d\omega' \int_{-\infty}^{\infty}\!\!\!d\omega \mathcal{T}_L(\omega' + \omega'')\mathcal{T}_R(\omega) \nonumber\\
 &\times\big[ f_R(\omega) \left(f_I(\omega + \omega'') -1 \right) f_I\left( \omega' + \omega'' \right) \left( f_L\left( \omega' \right) - 1 \right) \nonumber \\ 
&- \text{``}f \leftrightarrow (f-1)\text{''} \big] |\Xi_d(\omega'')|^2, 
\end{align}
where the emitter quantum dot transition  here fixes the energy $\omega' + \omega''$ in the reservoir region. At zero temperature, for $\delta$-like filters~(\ref{eq:deltaT}), and $\mu_I = \mu_R$, we find \\
\begin{align}
\label{eq:3bexplicit}
I_R^{\text{3b}} = &\frac{e}{2\pi} \frac{\Gamma^2}{4} \int_{\omega_L-\mu_L}^{\mu_I-\omega_R}\!\!\!\!\!\! d\omega'' \, |\Xi_d(\omega'')|^2 \nonumber \\
                                   &\times \left[ \theta(\mu_R- \omega_R) - \theta(\mu_L- \omega_R) \right]\nonumber \\
                                   &\quad \times \left[ \theta(\mu_I- \omega_L) - \theta(\mu_L + \mu_I - \omega_L- \omega_R) \right],
\end{align}
where
\begin{align}
 \Xi_d(\omega'') =  -\frac{1}{2\pi} \int_{-\infty}^{\infty} dq \, \nu_q^d \frac{\exp\left( i \frac{\Delta x}{v} \left(vq- \omega'' \right) \right)}{\left( vq-\omega'' -i\delta \right)^2}.
\end{align}
The second and third lines of~(\ref{eq:3bexplicit}) describe the outline of triangle \textbf{\rom{3}} in Fig.~\ref{fig:current}. For $d=0$,~(\ref{eq:3bexplicit}) constitutes the mirror image of~(\ref{eq:1bexplicit}) with respect to the $\omega_L$ coordinate, reflected at $(\mu_I+\mu_L)/2$.

\subsubsection{Inverted triangle hole current}

The third and final diagram generated by~(\ref{eq:HVIL}) is depicted in Fig.~\ref{fig:3c}, and the corresponding physical process is shown in Fig.~\ref{fig:3cProcess}. Here, the hole of the electron-hole pair generated by the virtual photon from the source enters the detector quantum dot.
The contribution of the diagram is given by\begin{align}
\label{eq:3cexplicit}
I_R^{\text{3c}} &= -\frac{e}{2\pi} \frac{\Gamma^2}{4} \int_{\mu_R-\omega_R}^{\mu_L-\omega_L}\!\!\!\!\!\! d\omega'' \, |\Xi_d(\omega'')|^2 \nonumber \\
                                   &\times \left[ \theta(\mu_L- \omega_L) - \theta(\mu_R - \omega_L) \right] \nonumber \\
                                   &\qquad\times  \left[ \theta(\mu_R- \omega_R) - \theta(\mu_R-\mu_L+\omega_L- \omega_R) \right].
\end{align}
The second and third line of~(\ref{eq:3cexplicit}) describe the outline of triangle \textbf{\rom{4}} observed in the measurement data of Fig.~\ref{fig:current}. Also in this case, for $d=0$ the contribution is the exact mirror image of the regular hole current~(\ref{eq:1cexplicit}), with respect to reflection of $\omega_L$ at $(\mu_I+\mu_L)/2$.

\section{Drain current for finite range model interaction}
\label{sec:GaussianModel}

In this section, all contributions  resulting from interaction Hamiltonians~(\ref{eq:HV}) and~(\ref{eq:HVIL}) are evaluated explicitly for the model interaction~(\ref{eq:GaussianModel}), previously employed in section~\ref{sec:CorrElastic}. In the source-reservoir interaction~(\ref{eq:HVIL}) the additional distance across the emitter quantum dot is phenomenologically accounted for  by the factor\footnote{The Fourier-transform of~(\ref{eq:nuq}) is proportional to $\exp{(-x/\lambda)}$.}
\begin{align}
\label{eq:interchannel}
\nu_q^d = \nu_q^{\vphantom{d}} \exp \left( -\frac{d}{\lambda} \right),
\end{align}
in which  $d$ corresponds to the spatial separation of source and reservoir channels. To obtain the current, the function $\Xi$, defined in~(\ref{eq:Xi}), has to be evaluated. Subsequently the energy integrals in~(\ref{eq:1bexplicit}), in~(\ref{eq:2bexplicit}), and in~(\ref{eq:2cexplicit}), have to be carried out to determine all inelastic contributions to the current. Evaluation and results for these integrals are presented in Appendix~\ref{app:EnergyIntegrals}.

\begin{figure}[t]
    \includegraphics[width=\textwidth]{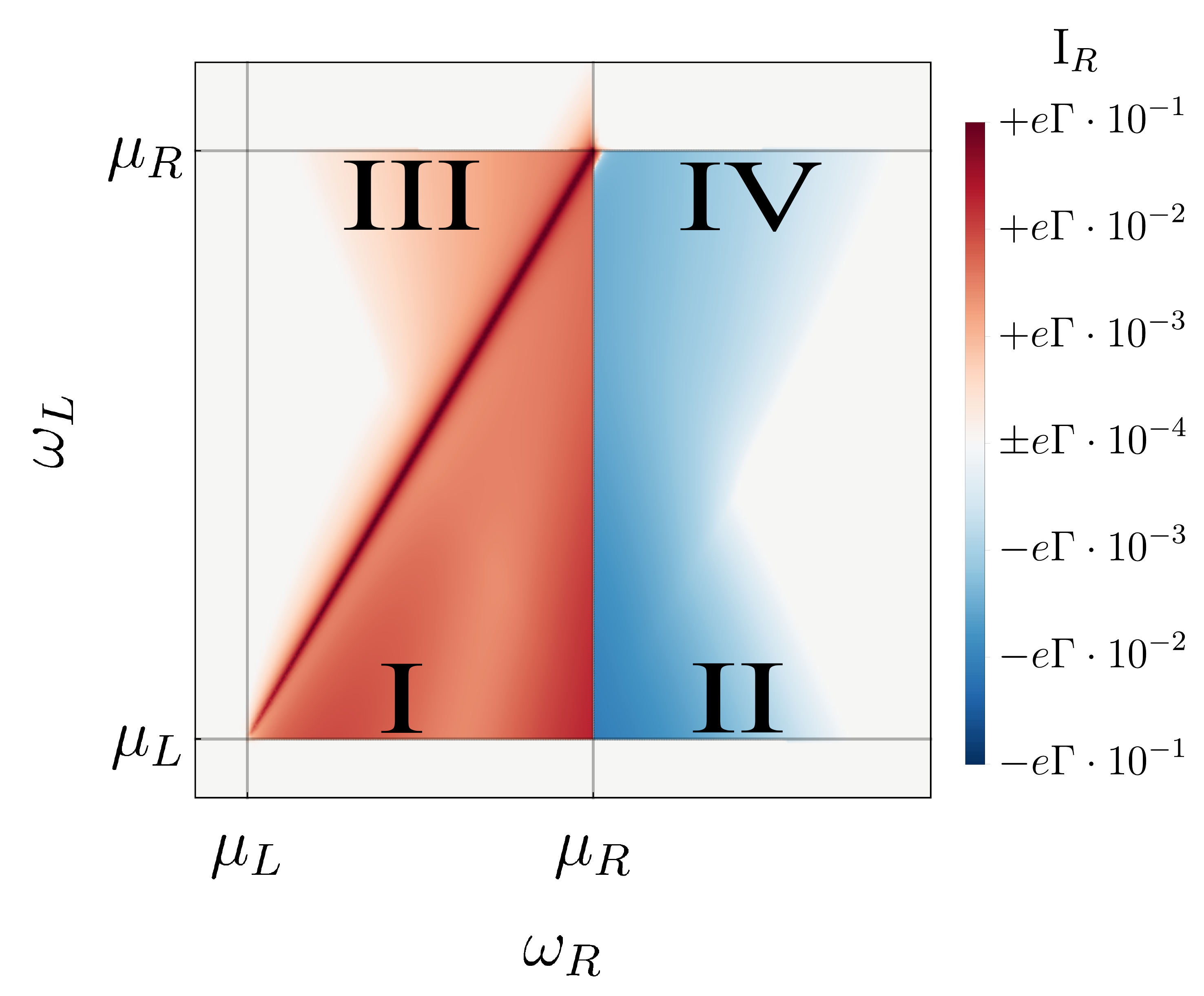}
    \caption{Drain current $\text{I}_R$ on logarithmic scale. The processes depicted in Figs.~\ref{fig:1aProcess},~\ref{fig:1bProcess}, and~\ref{fig:1cProcess}, caused by interactions in the reservoir region, generate current in the lower
 red and blue triangles (compare triangles \textbf{\rom{1}} and \textbf{\rom{2}} in Fig.~\ref{fig:current}). The processes of Figs.~\ref{fig:3bProcess} and~\ref{fig:3cProcess}, due to interactions between the reservoir region and the source lead, generate current in the inverted red and blue triangles (compare triangles \textbf{\rom{3}} and \textbf{\rom{4}} in Fig.~\ref{fig:current}). The present approach does not account for strong enough interactions   to completely suppress the elastic current, as seen in the experimental data. (Parameters are $\mu_L = \mu_R + 400\Gamma$, $v = 260 \lambda \Gamma$, $\nu_0 = 720\lambda \Gamma$, $\Delta x = 8\lambda$, $d =2.8\lambda$.) }
  \label{fig:TotalCurrent}
\end{figure}

In Fig.~\ref{fig:TotalCurrent}, the total drain current, accounting for all elastic and inelastic contributions presented in sections~\ref{sec:non-interactingI},~\ref{sec:interactions}, and~\ref{sec:HVIL}, is displayed on a logarithmic scale that 
preserves the current's sign. As anticipated in these sections, current is generated in all distinct regions in the space of detector and emitter energy in which the experiment shows a clear signal, compare Fig.~\ref{fig:current}. 
The contributions to the current in the individual triangles \textbf{\rom{1}}-\textbf{\rom{4}} in Figs.~\ref{fig:current} and~\ref{fig:TotalCurrent} are listed in Table~\ref{tab:list}.

\begin{table}[b]
\begin{tabular}{ c || c }
\label{tab:parameters}
  \textbf{triangle} &  \textbf{contribution (\#equation)}  \\
    \hline \hline
  \textbf{\rom{1}} & $ I_R^{\text{1a}}$~(\ref{eq:1aexplicit}), $I_R^{\text{1b}}$~(\ref{eq:1bexplicit}), $I_R^{\text{2a}}$~(\ref{eq:2aexplicit}), $I_R^{\text{2b}}$~(\ref{eq:2bexplicit}), $ I_R^{\text{3a}}$~(\ref{eq:3aexplicit})  \\
  \hline
 \textbf{\rom{2}} & $I_R^{\text{1c}}$~(\ref{eq:1cexplicit}), $I_R^{\text{2c}}$~(\ref{eq:2cexplicit}) \\
  \hline
    \textbf{\rom{3}} & $ I_R^{\text{3b}}$~(\ref{eq:3bexplicit}) \\
  \hline
  \textbf{\rom{4}} & $ I_R^{\text{3c}}$~(\ref{eq:3cexplicit}) \\
\end{tabular}
\caption{List of contributions in triangles \textbf{\rom{1}}-\textbf{\rom{4}}  to drain current in Figs.~\ref{fig:current} and~\ref{fig:TotalCurrent} due to inelastic processes.}
\label{tab:list}
\end{table}

While the reduction of the hole current (to the extent that the signal changes sign) along the hypotenuse of triangle~\textbf{\rom{2}} in the measurement data displayed in Fig.~\ref{fig:current} at $\omega_L - \mu_R \simeq \mu_R - \omega_R$ is mainly due to an excited state of the detector dot,~\cite{kraehenmann1} the reduction in Fig.~\ref{fig:TotalCurrent} at similar energies stems from the exchange diagram in Fig.~\ref{fig:2c}, reducing the current as a consequence of the indistinguishability of electrons. Higher order interaction terms in a model with several channels per edge, accounting for processes in which a reservoir electron excites an electron-hole pair in another channel, which in turn excites an electron-hole pair back in the reservoir, are not subject to such a suppression, 
since the exchange diagram does not appear for inter-channel interactions at the relevant tunneling order.

\begin{figure}[t]
    \includegraphics[width=.95\textwidth]{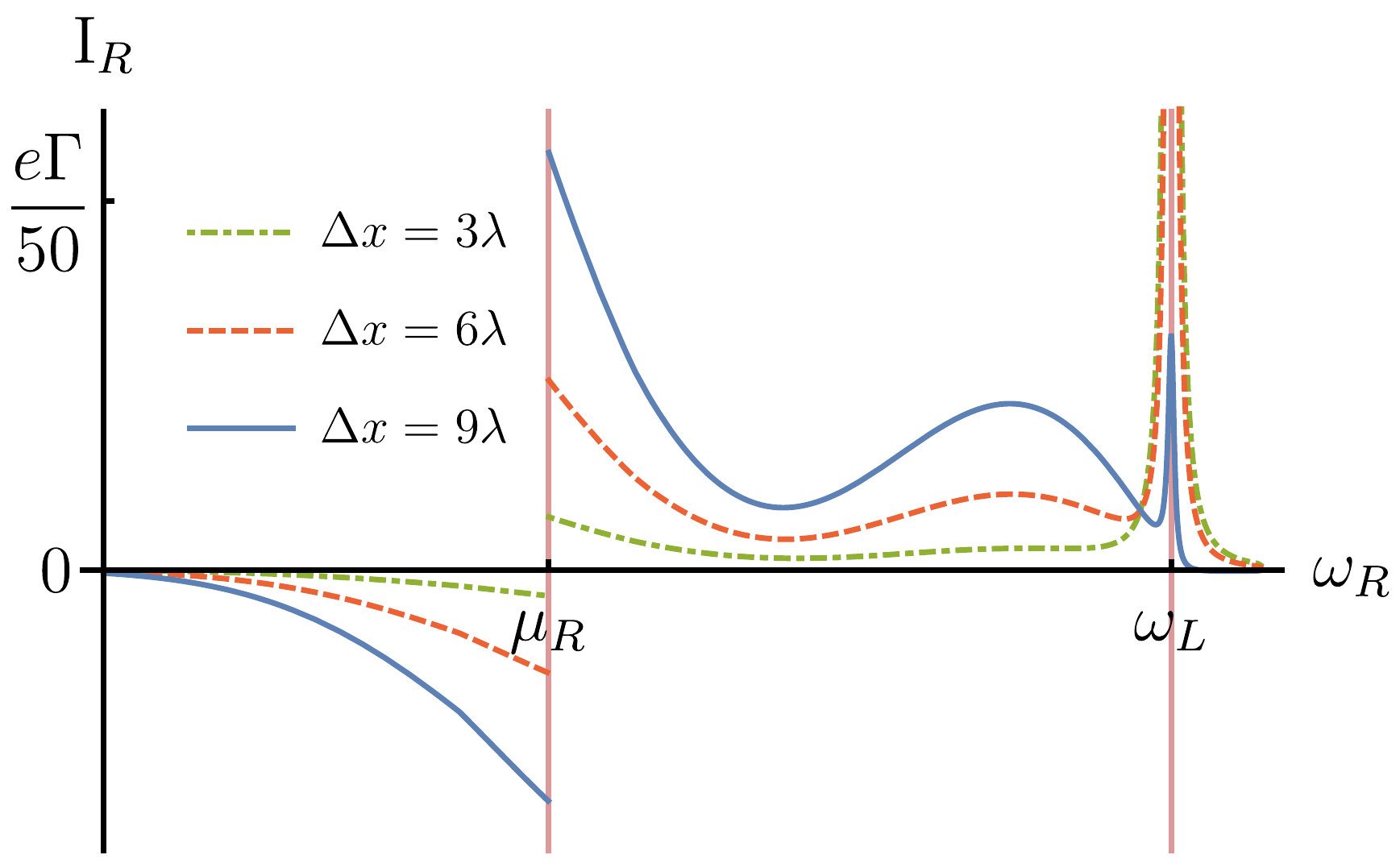}
        \caption{Detector current $\text{I}_R$ for increasing spatial separation $\Delta x$ between emitter and detector quantum dot. As the elastic peak  at $\omega_L=\omega_R$ diminishes for increasing separation, inelastic contributions intensify. The curve for the value $x=9\lambda$ (full line) is close to the
         boundary of validity of the approach, cf.~(\ref{eq:validity}), after which the elastic current takes on negative values and the approximation becomes unphysical. While the peak next to the elastic contribution is associated with processes in which the initial electron is directly transferred, see Fig.~\ref{fig:1aProcess}, the peaks close to the Fermi energy $\mu_R$ are associated with processes in which a charge carrier from the Fermi sea enters the detector, both for the regular triangles in Figs.~\ref{fig:1bProcess} and~\ref{fig:1cProcess}, as well as for the inverted triangles, Figs.~\ref{fig:3bProcess} and~\ref{fig:3cProcess}. (Parameters are $\mu_L = \mu_R + 400\Gamma$, $\omega_L = \mu_R + 350\Gamma$, $v = 260 \lambda \Gamma$, $\nu_0 = 720\lambda\Gamma$, $d =2.8\lambda$.) 
        }
  \label{fig:dynamics}
\end{figure}

In the experimental data, the elastic peak is not visible anymore already for comparably small dot energies. While the perturbative calculation to second order in interactions indicates  a diminishing elastic signal, the absence of the elastic line for higher filter energies escapes the approximation's admissible interaction strength.

Fig.~\ref{fig:dynamics} shows the drain current along a line cut in $\omega_R$ at constant $\omega_L$, for increasing dot separation $\Delta x$. With the separation also the transfer time $\Delta x / v$ increases, such that the sequence in Fig.~\ref{fig:dynamics} can be viewed as the temporal evolution of the electronic system in the channel (cf.~Ref.~[\onlinecite{PhysRevLett.113.166403}]). With the decrease of the elastic contribution, inelastic contributions intensify. While the process of Fig.~\ref{fig:1aProcess} leads to a current displaying a maximum next to the elastic peak at $\omega_L = \omega_R$, all remaining inelastic processes contribute close to and above (processes in Figs.~\ref{fig:1bProcess} and~\ref{fig:3bProcess}) or below (processes in Figs.~\ref{fig:1cProcess} and~\ref{fig:3cProcess}) the Fermi level. The reason for this localization in energy space is that a screened interaction, such as~(\ref{eq:GaussianModel}), is suppressed with increasing energy of virtual photons. While the photon energy (measured in units of $v/\lambda$) for the former process  is determined by the energy difference of the filters $\omega_L- \omega_R$, for the latter processes this energy depends on the distance of the filter energies from the Fermi level, $\omega_R-\mu_R$ and $\omega_L-\mu_R$, respectively.

\section{Summary and Outlook}
\label{sec:discussion}

A perturbative diagrammatic approach allows to relate individual signals measured in the ETH electron spectrometer to underlying physical processes. As a central point, the analysis shows that interactions between electrons in source and reservoir of the experimental sample generate current in regions of the detector-emitter energy landscape, in which the detector energy can exceed the emitter energy, thus giving rise to triangles \textbf{\rom{3}} and \textbf{\rom{4}} in the measurement data displayed in Fig.~\ref{fig:current}.
These currents are generated by Auger-like recombination processes in which the recombination energy is directly transferred from the source channel to the reservoir region. Thereby charge carriers are generated in the reservoir region at energies that can exceed the energy of electrons emitted from the source.

The experimental detection of energy transfer between leads and the central edge of the electron spectrometer, in combination with our theoretical analysis, suggests that such processes might have a significant impact on relaxation characteristics of mesoscopic devices, and that this decay channel cannot in general be readily neglected, as has previously been conjectured.\cite{PhysRevB.81.041311} It would be interesting to see in a quantitative study whether Auger-like processes can account for a significant amount of the energy loss reported in  Refs.~[\onlinecite{lesueur1}] and~[\onlinecite{Bocquillon2013}].  

For finite-range interactions, our approach furthermore demonstrates that processes, in which the original source electron enters the detector after dissipating some of its energy, contribute close to the elastic line, i.e.~at comparable dot energies. Processes in which the original electron is exchanged for a charge carrier from the Fermi sea, contribute close to the Fermi level. 

The perturbative nature of the approach in interactions limits its validity in terms of maximally admissible quantum dot energies (as measured from the Fermi energy), interaction strength, screening length, and spatial separation of  quantum dots (see discussion in the last paragraph of Section~\ref{sec:CorrElastic}), as well as number of interaction events. As a consequence, the present treatment is not suitable to quantitatively reproduce the signal measured in the sample, and to account for the absence of the elastic line in the measurement data at higher quantum dot energies. Relaxation due to inter-channel interaction between several channels in the reservoir edge cannot fully be captured by second order perturbation theory. Since this relaxation mechanism is not inhibited by Pauli blocking, it likely constitutes the dominant relaxation channel generating inelastic currents in triangles \textbf{\rom{1}} and \textbf{\rom{2}}. To extend the applicability of the formalism to the aforementioned scenarios, it is necessary to include higher order interaction terms.

A full-scale non-equilibrium bosonization approach,\cite{PhysRevLett.103.036801,gutman2,gutman4,gutman5,protopopov1,PhysRevB.85.075309,protopopov2} describing  interactions exactly, including also effects of interchannel interaction within the reservoir region, poses a particular challenge when tunneling through the dots and  finite range interactions are accounted for simultaneously. Our perturbative treatment indicates which elements have to enter an attempt for a full bosonization solution that captures currents in triangles \textbf{\rom{3}} and \textbf{\rom{4}}.

A further evident opportunity to apply the perturbative approach is the case in which the sample is not subject to an external magnetic field, such that electrons in the sample propagate in two dimensions. Here, the elastic line remains visible in the measurement data,~\cite{kraehenmann1} indicating that this case lies well within the validity of second order perturbation theory also at higher injection energies. Without the magnetic field, the geometry of the spectrometer plays a crucial role, with a significant enhancement of the signal at resonance energies of the sample. Within the general scheme of the present approach, such effects of geometry can be accommodated in the Green's functions that determine the transition amplitudes of the reservoir region.    

\begin{acknowledgments}
The authors would like to thank Igor Gornyi, Dmitry Polyakov and Bernd Rosenow, Klaus Ensslin, Thomas Ihn, Marc R\"o\"osli and Tobias Kr\"ahenmann,  as well as Kyrylo Snizhko and Tobias Holder, for useful discussions. Y.G.~acknowledges funding from DFG RO 2247/11-1, CRC 183 (project C01), and the Italia-Israel project QUANTRA. Y.M.~acknowledges support
 from ISF grant 292/15. J.P.~acknowledges support by the Koshland Foundation. S.G.F.~is a BGU PD fellow and acknowledges financial support from the Minerva and Kreitman foundations. 
\end{acknowledgments}

\appendix

\section{Derivation of general drain current  formula}
\label{app:DrainCurrent}

In this appendix, the intermediate steps which lead to the general expression~(\ref{eq:currentfinal}) for the drain current are presented.

Following~[\onlinecite{meir1}], the fact that the drain channel is assumed to be non-interacting allows us to expand
\begin{align}
\label{eq:GDD}
  \mathbf{G}_{\text{L}_\text{R} \text{D}_\text{R}} = \mathbf{g}_{\text{L}_\text{R} \text{L}_\text{R}} \mathbf{t}_{\text{L}_\text{R} \text{D}_\text{R}} \mathbf{G}_{\text{D}_\text{R} \text{D}_\text{R}}
\end{align}
in~(\ref{eq:currentinitial2}).
Here $\mathbf{g}_{\text{L}_\text{R} \text{L}_\text{R}}$ denotes the Green's function of the drain lead which develops solely with the bare Hamiltonian $H_0$, and $\mathbf{G}_{\text{D}_\text{R} \text{D}_\text{R}}$ denotes the Green's function of the detector quantum dot developing with the full Hamiltonian $H$. Expanding the Green's functions in~(\ref{eq:currentinitial2}) in their form~(\ref{eq:GDD}) on the Keldysh contour leads to the expression\cite{meir1,haug1}
\begin{align}
\label{eq:currentinitial3}
&I_R = \nonumber \\ &-\frac{e}{2 \pi} \int_{-\infty}^{\infty} d\omega \, \text{tr} \left\{ \mathbf{\Sigma}^<_{\text{D}_\text{R} \text{L}_\text{R} \text{D}_\text{R}} \mathbf{G}^>_{\text{D}_\text{R} \text{D}_\text{R}}   -  \mathbf{\Sigma}^>_{\text{D}_\text{R} \text{L}_\text{R} \text{D}_\text{R}} \mathbf{G}^<_{\text{D}_\text{R} \text{D}_\text{R}}    \right\}.
\end{align}
The Green's function of the detector quantum dot develops according to
\begin{align}
\label{eq:dotGreenfunctions}
  \mathbf{G}_{\text{D}_\text{R} \text{D}_\text{R}} = \mathbf{\overline{g}}_{\text{D}_\text{R} \text{D}_\text{R}} + \mathbf{\overline{g}}_{\text{D}_\text{R} \text{D}_\text{R}} \mathbf{\Sigma}_{\text{D}_\text{R} \text{I} \text{D}_\text{R}} \mathbf{\overline{g}}_{\text{D}_\text{R} \text{D}_\text{R}}.
\end{align} 
Expanding the lesser and greater Green's functions of the drain lead in (\ref{eq:currentinitial3}) in terms of (\ref{eq:dotGreenfunctions}), on the Keldysh contour, according to the Langreth rules,\cite{haug1}  we have for the lesser component
\begin{align}
\label{eq:dotGreenfunctionsKeldysh}
  &\mathbf{G}^<_{\text{D}_\text{R} \text{D}_\text{R}} = \mathbf{\overline{g}}^<_{\text{D}_\text{R} \text{D}_\text{R}} + \mathbf{\overline{g}}^<_{\text{D}_\text{R} \text{D}_\text{R}} \mathbf{\Sigma}^a_{\text{D}_\text{R} \text{I} \text{D}_\text{R}} \mathbf{\overline{g}}^a_{\text{D}_\text{R} \text{D}_\text{R}} \nonumber \\ &+ \mathbf{\overline{g}}^r_{\text{D}_\text{R} \text{D}_\text{R}} \mathbf{\Sigma}^<_{\text{D}_\text{R} \text{I} \text{D}_\text{R}} \mathbf{\overline{g}}^a_{\text{D}_\text{R} \text{D}_\text{R}} + \mathbf{\overline{g}}^r_{\text{D}_\text{R} \text{D}_\text{R}} \mathbf{\Sigma}^r_{\text{D}_\text{R} \text{I} \text{D}_\text{R}} \mathbf{\overline{g}}^<_{\text{D}_\text{R} \text{D}_\text{R}}.
\end{align}
The same relation follows for the greater component with ``$< \rightarrow >$''. After expansion of the lesser component of the Green's function~(\ref{eq:gdotleadR}) by means of the respective kinetic equation,~\cite{mahan2,haug1}
\begin{align}
\label{eq:gdotleadRkinetic}
  \mathbf{\overline{g}}^<_{\text{D}_\text{R} \text{D}_\text{R}} = \mathbf{\overline{g}}^r_{\text{D}_\text{R} \text{D}_\text{R}} \mathbf{\Sigma}^<_{\text{D}_\text{R} \text{L}_\text{R} \text{D}_\text{R}}  \mathbf{\overline{g}}^a_{\text{D}_\text{R} \text{D}_\text{R}},
\end{align}
 followed by subsequent insertion of~(\ref{eq:gdotleadRkinetic}) into~(\ref{eq:dotGreenfunctionsKeldysh}), only the third term of~(\ref{eq:dotGreenfunctionsKeldysh}) (and of the latter's counterpart for the greater Green's function) contributes in~(\ref{eq:currentinitial3}). To see this, the explicit expressions for the lesser and greater component of the drain lead's tunneling self-energy,
\begin{align}
  \mathbf{\Sigma}^<_{\text{D}_\text{R} \text{L}_\text{R} \text{D}_\text{R}} &= \sum_{k_R} t_{k_R R}^* g_{k_R}^<(\omega) t_{k_R R } \nonumber \\
  &\rightarrow \int_{-\infty}^{\infty} dk_R \rho \left( \omega_{k_R} \right) t_{k_R R}^* g_{k_R}^< (\omega) t_{k_R R } \nonumber \\
  &= i \Gamma(\omega) f_R(\omega),
\intertext{and}
  \mathbf{\Sigma}^>_{\text{D}_\text{R} \text{L}_\text{R} \text{D}_\text{R}} &=  i \Gamma(\omega)\left( f_R(\omega) - 1 \right),
\end{align}
respectively, have to be inserted into~(\ref{eq:currentinitial3}) and~(\ref{eq:gdotleadRkinetic}). Thereby, the first, second, and fourth term of~(\ref{eq:dotGreenfunctionsKeldysh}) cancel with the greater component counterpart   in~(\ref{eq:currentinitial3}), since the Fermi distributions of these terms are evaluated at the same chemical potential $\mu_R$ of the drain lead.  Insertion of the remaining third term of~(\ref{eq:dotGreenfunctionsKeldysh}) into~(\ref{eq:currentinitial3}) then leads to the desired general expression for the current~(\ref{eq:currentfinal}).

Using the kinetic equations for the retarded and advanced components of the dot-lead Green's functions
\begin{align}
\label{eq:gdotleadRretardedadvanced}
  \mathbf{\overline{g}}^{r/a}_{\text{D}_\text{R} \text{D}_\text{R}} = \mathbf{g}^{r/a}_{\text{D}_\text{R} \text{D}_\text{R}} + \mathbf{g}^{r/a}_{\text{D}_\text{R} \text{D}_\text{R}} \mathbf{\Sigma}^{r/a}_{\text{D}_\text{R} \text{L}_\text{R} \text{D}_\text{R}}  \mathbf{\overline{g}}^{r/a}_{\text{D}_\text{R} \text{D}_\text{R}},
\end{align}
and insertion of the explicit expressions for the lead self-energies in the wide band approximation,\cite{meir1,haug1}
\begin{align}
  \mathbf{\Sigma}^{r/a}_{\text{D}_\text{R} \text{L}_\text{R} \text{D}_\text{R}} \rightarrow \mp \frac{i}{2} \Gamma(\omega),
\end{align}
allows to cast the general formula~(\ref{eq:currentfinal}) for the current into the explicit form~(\ref{eq:currentfinalexplicit}).

\section{Derivation of drain current in absence of interactions}
\label{app:CurrentNoV}

In this appendix, evaluation of expressions which determine the current~(\ref{eq:currentexplicit}) in the absence of interactions are presented.

Upon insertion of~(\ref{eq:GTkinetic}) into~(\ref{eq:currentfinal}), determination of the drain current requires evaluation of the following transition self-energies: the transition self-energy from the right dot to the left dot vanishes,
\begin{align}
    \mathbf{t}_{\text{D}_\text{L} \text{I}} \mathbf{G}^{Tr}_{II} \mathbf{t}_{\text{I} \text{D}_\text{R}} &= 0,
\end{align}
since the channel is chiral with right-moving particles only. For the transition from left to right, we find the Dyson equation
\begin{align}
\label{eq:chiralDyson}
  \mathbf{t}_{\text{D}_\text{R} \text{I}} \mathbf{G}^{Tr}_{II} \mathbf{t}_{\text{I} \text{D}_\text{L}} &=  \mathbf{t}_{\text{D}_\text{R} \text{I}} \mathbf{g}^{r}_{II} \mathbf{t}_{\text{I} \text{D}_\text{L}} + \mathbf{t}_{\text{D}_\text{R} \text{I}} \mathbf{g}^{r}_{II} \mathbf{\Sigma}^r_{\text{I}\text{D}_{\text{R}}\text{I}} \mathbf{G}^{Tr}_{II} \mathbf{t}_{\text{I} \text{D}_\text{L}} \nonumber \\
  &\qquad \qquad \qquad \,\,\,  + \mathbf{t}_{\text{D}_\text{R} \text{I}} \mathbf{g}^{r}_{II} \mathbf{\Sigma}^r_{\text{I}\text{D}_{\text{L}}\text{I}} \mathbf{\overline{g}}^{r}_{II} \mathbf{t}_{\text{I} \text{D}_\text{L}},
  \end{align}
where
\begin{align}
  \mathbf{\Sigma}_{\text{I}\text{D}_{\text{L/R}}\text{I}} &= \mathbf{t}_{\text{I} \text{D}_\text{L/R}} \mathbf{\overline{g}}_{\text{D}_\text{L/R} \text{D}_\text{L/R}} \mathbf{t}_{\text{D}_\text{L/R} \text{I}},
  \intertext{with}
  \label{eq:localDyson}
    \mathbf{\overline{g}}_{\text{II}} &=  \mathbf{g}_{II} + \mathbf{g}_{II} \mathbf{\Sigma}_{\text{I} \text{D}_\text{L} \text{I}}  \mathbf{\overline{g}}_{\text{II}}.
\end{align}
The Dyson equation~(\ref{eq:chiralDyson}) reflects that charge carriers can only pass from the left dot to the right dot once, before and after tunneling back and forth between the reservoir and each of  the dots. For the local self-energy at the left dot, we find
\begin{align}
  \mathbf{t}_{\text{D}_\text{L} \text{I}} \mathbf{G}^{Tr}_{II} \mathbf{t}_{\text{I} \text{D}_\text{L}} &= \mathbf{t}_{\text{D}_\text{L} \text{I}} \mathbf{\overline{g}}^{r}_{II} \mathbf{t}_{\text{I} \text{D}_\text{L}}.
\end{align}
An analogous equation holds for the local self-energy at the right dot, since going back and forth between the left and the right dot is prohibited in the chiral channel. The corresponding relations for the advanced Green's functions are obtained by complex conjugation.

The final building block to obtain an explicit expression for the current are the transition self-energies for the bare Green's functions of the intermediate region, which are given by 
\begin{align}
\label{eq:transitiona}
   \mathbf{t}_{\text{D}_\text{L} \text{I}} \mathbf{g}^{r}_{II} \mathbf{t}_{\text{I} \text{D}_\text{R}} &= 0, \nonumber \\
   \mathbf{t}_{\text{D}_\text{R} \text{I}} \mathbf{g}^{r}_{II} \mathbf{t}_{\text{I} \text{D}_\text{L}} &= -i\Gamma e^{ +i \frac{\Delta x}{v} \omega}, \nonumber \\
   \mathbf{t}_{\text{D}_\text{L} \text{I}} \mathbf{g}^{r}_{II} \mathbf{t}_{\text{I} \text{D}_\text{L}} &= -\frac{i\Gamma}{2}, \nonumber \\
   \mathbf{t}_{\text{D}_\text{R} \text{I}} \mathbf{g}^{r}_{II} \mathbf{t}_{\text{I} \text{D}_\text{R}} &= -\frac{i\Gamma}{2},
\end{align}
where also here the advanced components are obtained by complex conjugation, as well as by
\begin{align}
\label{eq:transitionlesser}
  \mathbf{t}_{\text{D}_\text{L} \text{I}} \mathbf{g}^{<}_{II} \mathbf{t}_{\text{I} \text{D}_\text{R}} &= i\Gamma f_I(\omega) e^{ -i \frac{\Delta x}{v} \omega}, \nonumber \\
  \mathbf{t}_{\text{D}_\text{R} \text{I}} \mathbf{g}^{<}_{II} \mathbf{t}_{\text{I} \text{D}_\text{L}} &= i\Gamma f_I(\omega) e^{ +i \frac{\Delta x}{v} \omega}, \nonumber \\
  \mathbf{t}_{\text{D}_\text{L} \text{I}} \mathbf{g}^{<}_{II} \mathbf{t}_{\text{I} \text{D}_\text{L}} &= i\Gamma f_I(\omega), \nonumber \\
  \mathbf{t}_{\text{D}_\text{R} \text{I}} \mathbf{g}^{<}_{II} \mathbf{t}_{\text{I} \text{D}_\text{R}} &= i\Gamma f_I(\omega),
\end{align}
where $\Delta x = x_R- x_L >0$. The respective greater Green's functions are obtained by the replacement ``$f \rightarrow (f - 1)$''.

Solving above Dyson equations~(\ref{eq:chiralDyson}) and~(\ref{eq:localDyson}), and insertion of the explicit transition self-energies~(\ref{eq:transitiona}) and~(\ref{eq:transitionlesser}) into~(\ref{eq:currentfinal}), leads to the explicit current~(\ref{eq:currentexplicit}) of the non-interacting system.

\section{Contributions due to reservoir electron interactions}

In this appendix, Green's function expressions which give rise to changes of the elastic current as well as to inelastic contributions due to reservoir electron interactions are collected. 

\subsection{Green's function expressions for corrections to elastic current }
\label{app:ElasticCorrection}

Insertion of the second and fourth term of expansion~(\ref{eq:altGLangreth}) into the general expression for the current~(\ref{eq:currentfinal}), and expansion of the lesser/greater Green's functions of the reservoir region~(\ref{eq:GTkinetic}), leads to
\begin{widetext}
\begin{align}
\label{eq:currentcorrection}
I_R^{\text{corr}} =  -\frac{e}{2 \pi} \int_{-\infty}^{\infty} d\omega \, \bigg[ \text{tr} \big\{ &\mathbf{\overline{g}}^a_{\text{D}_\text{R} \text{D}_\text{R}} \mathbf{\Sigma}^<_{\text{D}_\text{R} \text{L}_{\text{R}} \text{D}_\text{R}} \mathbf{\overline{g}}^r_{\text{D}_\text{R} \text{D}_\text{R}} \textbf{t}_{\text{D}_\text{R}\text{I}} \mathbf{G}^{Tr}_{\text{II}} \mathbf{t}_{\text{I}\text{D}_\text{L}}\mathbf{\overline{g}}^r_{\text{D}_\text{L} \text{D}_\text{L}} \mathbf{\Sigma}^>_{\text{D}_\text{L} \text{L}_{\text{L}} \text{D}_\text{L}} \mathbf{\overline{g}}^a_{\text{D}_\text{L} \text{D}_\text{L}}\mathbf{t}_{\text{D}_\text{L}\text{I}} \mathbf{G}^{Ta}_{\text{II}} \mathbf{\Sigma}^a_{\text{IVI}} \mathbf{G}^{Ta}_{\text{II}} \textbf{t}_{\text{I}\text{D}_\text{R}} \nonumber \\ 
+&\mathbf{\overline{g}}^a_{\text{D}_\text{R} \text{D}_\text{R}} \mathbf{\Sigma}^<_{\text{D}_\text{R} \text{L}_{\text{R}} \text{D}_\text{R}} \mathbf{\overline{g}}^r_{\text{D}_\text{R} \text{D}_\text{R}}  \textbf{t}_{\text{D}_\text{R}\text{I}}\mathbf{G}^{Tr}_{\text{II}} \mathbf{\Sigma}^r_{\text{IVI}} \mathbf{G}^{Tr}_{\text{II}} \mathbf{t}_{\text{I}\text{D}_\text{L}}\mathbf{\overline{g}}^r_{\text{D}_\text{L} \text{D}_\text{L}} \mathbf{\Sigma}^>_{\text{D}_\text{L} \text{L}_{\text{L}} \text{D}_\text{L}} \mathbf{\overline{g}}^a_{\text{D}_\text{L} \text{D}_\text{L}}\mathbf{t}_{\text{D}_\text{L}\text{I}} \mathbf{G}^{Ta}_{\text{II}}\textbf{t}_{\text{I}\text{D}_\text{R}} \big\} 
 -  \text{``}< \leftrightarrow >\text{''}   \bigg].
\end{align}
\end{widetext}
For the model interaction~(\ref{eq:GaussianModel}), we obtain for the polarization diagram term~(\ref{eq:Epol}) the contribution
\begin{align}
\label{eq:ePolExplicit}
  E^2_{\text{pol}}(\omega) = \frac{1}{(2\pi)^2} \frac{\nu_0^2}{2\lambda^2} \frac{\frac{\lambda^2}{v^2}\left( \mu_I - \omega \right)^2}{1+ \frac{\lambda^2}{v^2}\left( \mu_I - \omega \right)^2} ,
\end{align}
and for the exchange diagram term~(\ref{eq:Eexch}) the contribution
\begin{align}
\label{eq:eExchExplicit}
  E^2_{\text{exch}}(\omega) &= \frac{1}{(2\pi)^2} \frac{\nu_0^2}{\lambda^2} \bigg[ \arctan^2 \left( \frac{\lambda}{v} \left( \mu_I-\omega \right) \right) \nonumber \\ &
   \qquad\qquad\qquad\qquad- h\left( \frac{\lambda}{v} \left( \mu_I-\omega \right) \right) \bigg],
\end{align}
in which
\begin{align}
h(x) = x \int_{0}^{1} dy \frac{\arctan (xy)}{1+x^2(1-y)^2}.
\end{align}
The poles of~(\ref{eq:ePolExplicit}) are located at $\omega =  \mu \pm i v/\lambda$ and $E^2_{\text{pol}}(\omega \to \infty) \to \nu_0^2 / 2 (2\pi\lambda)^2$ in the complex plane. A numerical comparison shows that~(\ref{eq:eExchExplicit}) has a similar pole structure as~(\ref{eq:ePolExplicit}) and $E^2_{\text{exch}}(\omega \to \infty) \to 0$ in the complex plane.

In~(\ref{eq:currentcorrection}), the poles of~(\ref{eq:ePolExplicit}) contribute terms that decay as $\exp \left( - \Delta x / \lambda \right)$. Due to the similarity of the pole structure of~(\ref{eq:eExchExplicit}) we expect a similar decay also in the exchange contribution.

Neglecting the poles of~(\ref{eq:ePolExplicit}) and~(\ref{eq:eExchExplicit}), the corrected transition self-energies in~(\ref{eq:currentcorrection}) are given by 
 \begin{align}
 \label{eq:ElasticCorrContour}
   \textbf{t}_{\text{D}_\text{R}\text{I}}\mathbf{G}^{Tr}_{\text{II}} \mathbf{\Sigma}^r_{\text{IVI}} \mathbf{G}^{Tr}_{\text{II}} \mathbf{t}_{\text{I}\text{D}_\text{L}} &\simeq 
   \textbf{t}_{\text{D}_\text{R}\text{I}}\mathbf{g}^{r}_{\text{II}} \left[\mathbf{g}^{r}_{\text{II}} E^2(vk) \right] \mathbf{g}^{r}_{\text{II}} \mathbf{t}_{\text{I}\text{D}_\text{L}} \nonumber \\
   &\simeq  \frac{i\Gamma}{2} \left. \frac{\partial^2}{\partial k^2} \left[ \exp\left(i \Delta x k \right)E(vk)  \right] \right|_{k = \frac{\omega}{v}}.
 \end{align}
 According to the above considerations, the approximation in the second line of~(\ref{eq:ElasticCorrContour}) is justified as long as $\lambda \ll \Delta x$.  Combining all terms in~(\ref{eq:currentcorrection}) then generates the second and third term in~(\ref{eq:correlastic}). We neglect the third term, which is suppressed against the second term by a factor of $\lambda^2/\Delta x^2$.
 
 For the evaluation of the $\omega$ integral in~(\ref{eq:correlastic}) we take into account residues of $\mathcal{T}_{L/R}$ and not of $E^2$. For~(\ref{eq:ePolExplicit}), the contribution of the latter is suppressed by a factor of $2(1/2\pi)^3(\nu_0/v)^2 (\Delta x / \lambda)^2 (\Gamma\lambda/v)^3$ against the maximum of the elastic current, given $\mu_L-\mu_R \sim v/\lambda$.

 \subsection{Inelastic contributions in terms of Green's functions}
\label{app:InelasticContributions}
 
 After collection of all tunneling terms which correspond to one electron or hole transmission per quantum dot, and discarding the remaining terms, no tunneling amplitudes remain in the reservoir equilibration term  in~(\ref{eq:GTkineticelaborate}),
\begin{align}
\label{eq:noreflections}
  \left(\mathbf{I} + \mathbf{G}^{Tr}_{\text{II}}  \mathbf{\Sigma}^r_\text{ITI} \right) \mathbf{g}^{</>}_{\text{II}} \left(\mathbf{I} + \mathbf{\Sigma}^a_\text{ITI} \mathbf{G}^{Ta}_{\text{II}}  \right) \rightarrow \mathbf{g}^{</>}_{\text{II}}.
\end{align}
 After the initial approximation~(\ref{eq:noreflections}), the contributions due to the diagrams in Figs.~\ref{fig:1a},~\ref{fig:1b},~\ref{fig:1c},~\ref{fig:2a}, and~\ref{fig:2c} are, respectively, given by
\begin{widetext}
\begin{align}
\label{eq:I1a}
 I_R^{\text{1a}} = -\frac{e}{2\pi} \int_{-\infty}^{\infty}\!\!\! d\omega''\, \nu_{\text{Q}}^2 \bigg[ &\frac{1}{2\pi\Omega} \int_{-\infty}^{\infty}\!\!\! d\omega' \, \text{tr}\left\{ \mathbf{g}^{<}_{\text{I+QI+Q}}(\omega' +\omega'') \mathbf{g}^{>}_{\text{II}}(\omega') \right\} \nonumber \\
 \times &\frac{1}{2\pi\Omega} \int_{-\infty}^{\infty}\!\!\! d\omega \, \text{tr}\big\{ 
 \mathbf{\overline{g}}^a_{\text{D}_\text{R} \text{D}_\text{R}}(\omega) \mathbf{\Sigma}^<_{\text{D}_\text{R} \text{L}_{\text{R}} \text{D}_\text{R}}(\omega) \mathbf{\overline{g}}^r_{\text{D}_\text{R} \text{D}_\text{R}}(\omega)
 \textbf{t}_{\text{D}_\text{R}\text{I}} \mathbf{G}^{Tr}_{\text{II}}(\omega) \mathbf{G}^{Tr}_{\text{I+QI}}(\omega+\omega'') \mathbf{t}_{\text{I}\text{D}_\text{L}} \nonumber \\
\times &\mathbf{\overline{g}}^r_{\text{D}_\text{L} \text{D}_\text{L}}(\omega+\omega'') \mathbf{\Sigma}^>_{\text{D}_\text{L} \text{L}_{\text{L}} \text{D}_\text{L}}(\omega+\omega'') \mathbf{\overline{g}}^a_{\text{D}_\text{L} \text{D}_\text{L}}(\omega+\omega'')
 \mathbf{t}_{\text{D}_\text{L}\text{I}} \mathbf{G}^{Ta}_{\text{II+Q}}(\omega+\omega'')\mathbf{G}^{Ta}_{\text{II}}(\omega)\textbf{t}_{\text{I}\text{D}_\text{R}} \big\}  -  \text{``}< \leftrightarrow >\text{''}  
\bigg],
\end{align}

\begin{align}
\label{eq:I1b}
I_R^{\text{1b}} = -\frac{e}{2\pi} \int_{-\infty}^{\infty}\!\!\!d\omega'' \, \nu_Q \nu_{Q'} \bigg[
   \frac{1}{2\pi\Omega}\int_{-\infty}^{\infty}\!\!\! d\omega' \, \text{tr}\big\{
&\mathbf{\overline{g}}^r_{\text{D}_\text{L} \text{D}_\text{L}}(\omega') \mathbf{\Sigma}^>_{\text{D}_\text{L} \text{L}_{\text{L}} \text{D}_\text{L}}(\omega') \mathbf{\overline{g}}^a_{\text{D}_\text{L} \text{D}_\text{L}}(\omega') \nonumber \\
&\textbf{t}_{\text{D}_\text{L}\text{I}} \mathbf{G}^{Ta}_{\text{II+Q}}(\omega') \mathbf{g}^{<}_{\text{I+Q+Q'I+Q+Q'}}(\omega' + \omega'') \mathbf{G}^{Tr}_{\text{I+Q'I}}(\omega') \mathbf{t}_{\text{I}\text{D}_\text{L}}
  \big\} \nonumber \\
 \times \frac{1}{2\pi\Omega}\int_{-\infty}^{\infty}\!\!\! d\omega \, \text{tr}\big\{
&\mathbf{\overline{g}}^a_{\text{D}_\text{R} \text{D}_\text{R}}(\omega) \mathbf{\Sigma}^<_{\text{D}_\text{R} \text{L}_{\text{R}} \text{D}_\text{R}}(\omega) \mathbf{\overline{g}}^r_{\text{D}_\text{R} \text{D}_\text{R}}(\omega) \nonumber \\
&\textbf{t}_{\text{D}_\text{R}\text{I}} \mathbf{G}^{Tr}_{\text{II+Q'}}(\omega) \mathbf{g}^{>}_{\text{I+Q+Q'I+Q+Q'}}(\omega + \omega'') \mathbf{G}^{Ta}_{\text{I+QI}}(\omega) \mathbf{t}_{\text{I}\text{D}_\text{R}}
  \big\}  
  -  \text{``}< \leftrightarrow >\text{''}  
\bigg],
\end{align}

\begin{align}
\label{eq:I1c}
I_R^{\text{1c}} = -\frac{e}{2\pi} \int_{-\infty}^{\infty}\!\!\!d\omega'' \, \nu_Q \nu_{Q'} \bigg[
   \frac{1}{2\pi\Omega}\int_{-\infty}^{\infty}\!\!\! d\omega' \, \text{tr}\big\{
&\mathbf{\overline{g}}^r_{\text{D}_\text{L} \text{D}_\text{L}}(\omega') \mathbf{\Sigma}^<_{\text{D}_\text{L} \text{L}_{\text{L}} \text{D}_\text{L}}(\omega') \mathbf{\overline{g}}^a_{\text{D}_\text{L} \text{D}_\text{L}}(\omega') \nonumber \\
&\textbf{t}_{\text{D}_\text{L}\text{I}} \mathbf{G}^{Ta}_{\text{II+Q}}(\omega') \mathbf{g}^{>}_{\text{II}}(\omega' + \omega'') \mathbf{G}^{Tr}_{\text{I+Q'I}}(\omega') \mathbf{t}_{\text{I}\text{D}_\text{L}}
  \big\} \nonumber \\
 \times \frac{1}{2\pi\Omega}\int_{-\infty}^{\infty}\!\!\! d\omega \, \text{tr}\big\{
&\mathbf{\overline{g}}^a_{\text{D}_\text{R} \text{D}_\text{R}}(\omega) \mathbf{\Sigma}^<_{\text{D}_\text{R} \text{L}_{\text{R}} \text{D}_\text{R}}(\omega) \mathbf{\overline{g}}^r_{\text{D}_\text{R} \text{D}_\text{R}}(\omega) \nonumber \\
&\textbf{t}_{\text{D}_\text{R}\text{I}} \mathbf{G}^{Tr}_{\text{II+Q'}}(\omega) \mathbf{g}^{>}_{\text{I+Q+Q'I+Q+Q'}}(\omega + \omega'') \mathbf{G}^{Ta}_{\text{I+QI}}(\omega) \mathbf{t}_{\text{I}\text{D}_\text{R}}
  \big\}  
  -  \text{``}< \leftrightarrow >\text{''}  
\bigg],
\end{align}

\begin{align}
\label{eq:I2a}
I_R^{\text{2a}} = \frac{e}{2\pi} \nu_Q& \nu_{Q'} \int_{-\infty}^{\infty}\!\!\!d\omega'' \, \int_{-\infty}^{\infty}\!\!\! d\omega' \, \int_{-\infty}^{\infty}\!\!\! d\omega \,     \frac{1}{(2\pi\Omega)^2} \bigg[
 \text{tr}\big\{
 \mathbf{\overline{g}}^a_{\text{D}_\text{R} \text{D}_\text{R}}(\omega) \mathbf{\Sigma}^<_{\text{D}_\text{R} \text{L}_{\text{R}} \text{D}_\text{R}}(\omega) \mathbf{\overline{g}}^r_{\text{D}_\text{R} \text{D}_\text{R}}(\omega) \nonumber \\
 \times &\textbf{t}_{\text{D}_\text{R}\text{I}} \mathbf{G}^{Tr}_{\text{II}}(\omega) \mathbf{G}^{Tr}_{\text{I+QI}}(\omega+\omega'') \mathbf{t}_{\text{I}\text{D}_\text{L}}  \mathbf{\overline{g}}^r_{\text{D}_\text{L} \text{D}_\text{L}}(\omega + \omega'' ) \mathbf{\Sigma}^>_{\text{D}_\text{L} \text{L}_{\text{L}} \text{D}_\text{L}}(\omega + \omega'' ) \mathbf{\overline{g}}^a_{\text{D}_\text{L} \text{D}_\text{L}}(\omega + \omega'' ) \nonumber \\
\times &\textbf{t}_{\text{D}_\text{L}\text{I}} \mathbf{G}^{Ta}_{\text{II+Q}}(\omega+\omega'') \mathbf{g}^{<}_{\text{I+Q+Q'I+Q+Q'}}( \omega+ \omega' + \omega'') \mathbf{g}^{>}_{\text{I+Q'I+Q'}}(\omega + \omega') \mathbf{G}^{Ta}_{\text{II}}(\omega) \textbf{t}_{\text{I}\text{D}_\text{R}} 
  \big\}  
  -  \text{``}< \leftrightarrow >\text{''}  
\bigg],
\end{align}

\begin{align}
\label{eq:I2c}
I_R^{\text{2c}} = \frac{e}{2\pi} \nu_Q \nu_{Q'} &\int_{-\infty}^{\infty}\!\!\!d\omega'' \, \int_{-\infty}^{\infty}\!\!\! d\omega' \, \int_{-\infty}^{\infty}\!\!\! d\omega \, \nonumber \\
    \times\frac{1}{(2\pi\Omega)^2} \bigg[
 \text{tr}\big\{
 &\mathbf{\overline{g}}^a_{\text{D}_\text{R} \text{D}_\text{R}}(\omega) \mathbf{\Sigma}^<_{\text{D}_\text{R} \text{L}_{\text{R}} \text{D}_\text{R}}(\omega) \mathbf{\overline{g}}^r_{\text{D}_\text{R} \text{D}_\text{R}}(\omega)
 \textbf{t}_{\text{D}_\text{R}\text{I}} \mathbf{G}^{Tr}_{\text{II}}(\omega) \mathbf{g}^{>}_{\text{I+QI+Q}}(\omega+\omega'') \mathbf{G}^{Tr}_{\text{I+Q+Q'I}}(\omega+\omega'+\omega'') \mathbf{t}_{\text{I}\text{D}_\text{L}}   \nonumber \\
 \times &\mathbf{\overline{g}}^r_{\text{D}_\text{L} \text{D}_\text{L}}(\omega+\omega'+\omega'') \mathbf{\Sigma}^<_{\text{D}_\text{L} \text{L}_{\text{L}} \text{D}_\text{L}}(\omega+\omega'+\omega'') \mathbf{\overline{g}}^a_{\text{D}_\text{L} \text{D}_\text{L}}(\omega+\omega'+\omega'') \nonumber \\
\times&\textbf{t}_{\text{D}_\text{L}\text{I}} \mathbf{G}^{Ta}_{\text{II+Q+Q'}}(\omega+ \omega'+ \omega'') \mathbf{g}^{<}_{\text{I+Q'I+Q'}}( \omega+ \omega' ) \mathbf{G}^{Ta}_{\text{II}}(\omega) \textbf{t}_{\text{I}\text{D}_\text{R}} 
  \big\}  
  -  \text{``}< \leftrightarrow >\text{''}  
\bigg].
\end{align}
\end{widetext}
The subscript Q in the index of the reservoir Green's functions indicates that the photon momentum $q$ has to be added to the index and that the index is summed over. While the first trace in the polarization diagram expressions~(\ref{eq:I1a}),~(\ref{eq:I1b}), and~(\ref{eq:I1c}) corresponds to the closed Fermion loop, the second trace in the same expressions corresponds to the lower line of the respective diagrams.

\section{Tunneling Green's functions of  source channel}
\label{app:sourceGF}

In this appendix, the source channel tunneling Green's functions are presented, which are required to determine the diagrams that contribute due to interchannel interaction~(\ref{eq:HVIL}).

These Green's functions develop according to
\begin{align}
\label{eq:GTL}
  \mathbf{G}^T_{\text{L}_\text{L}\text{L}_\text{L}} = \mathbf{g}_{\text{L}_\text{L}\text{L}_\text{L}}+  \mathbf{g}_{\text{L}_\text{L}\text{L}_\text{L}} \mathbf{\Sigma}_{\text{L}_\text{L}\text{D}_\text{L}\text{L}_\text{L}} \mathbf{g}_{\text{L}_\text{L}\text{L}_\text{L}},
\end{align}
where
\begin{align}
\label{eq:appD}
\mathbf{\Sigma}_{\text{L}_\text{L}\text{D}_\text{L}\text{L}_\text{L}} = \mathbf{t}_{\text{L}_\text{L}\text{D}_\text{L}} \mathbf{G}^T_{\text{D}_\text{L}\text{D}_\text{L}} \mathbf{t}_{\text{D}_\text{L}\text{L}_\text{L}},
\end{align}
in which
\begin{align}
\label{eq:dotLGreenfunctions}
  \mathbf{G}^{T}_{\text{D}_\text{L} \text{D}_\text{L}} = \mathbf{\overline{g}}_{\text{D}_\text{L} \text{D}_\text{L}} + \mathbf{\overline{g}}_{\text{D}_\text{L} \text{D}_\text{L}}\mathbf{t}_{\text{D}_\text{L}\text{I}} \mathbf{G}^T_{\text{II}} \mathbf{t}_{\text{I}\text{D}_\text{L}} \mathbf{\overline{g}}_{\text{D}_\text{L} \text{D}_\text{L}}.
\end{align}
Upon insertion of~(\ref{eq:appD}) into~(\ref{eq:GTL}), application of the Langreth rules, and subsequent collection of all terms, the lesser/greater component of~(\ref{eq:GTL}) on the Keldysh contour becomes
\begin{scriptsize}
\begin{align}
\label{eq:GTLlesser}
  \mathbf{G}^{T{</>}}_{\text{L}_\text{L}\text{L}_\text{L}} = \left( \mathbf{I} + \mathbf{g}^r_{\text{L}_\text{L}\text{L}_\text{L}}\mathbf{\Sigma}^r_{\text{L}_\text{L}\text{D}_\text{L}\text{L}_\text{L}}\right) &\mathbf{g}^{</>}_{\text{L}_\text{L}\text{L}_\text{L}}\left( \mathbf{I} + \mathbf{\Sigma}^a_{\text{L}_\text{L}\text{D}_\text{L}\text{L}_\text{L}} \mathbf{g}^a_{\text{L}_\text{L}\text{L}_\text{L}}\right) \nonumber \\
  + \mathbf{g}^r_{\text{L}_\text{L}\text{L}_\text{L}} \mathbf{\overline{\Sigma}}^r_{\text{L}_\text{L}\text{D}_\text{L}\text{I}} \left(\mathbf{I} + \mathbf{G}^{Tr}_{\text{II}}  \mathbf{\Sigma}^r_\text{ITI} \right) &\mathbf{g}^{{</>}}_{\text{II}} \left(\mathbf{I} + \mathbf{\Sigma}^a_\text{ITI} \mathbf{G}^{Ta}_{\text{II}}  \right) \mathbf{\overline{\Sigma}}^a_{\text{I}\text{D}_\text{L}\text{L}_\text{L}} \mathbf{g}^a_{\text{L}_\text{L}\text{L}_\text{L}} \nonumber \\
  + \mathbf{g}^r_{\text{L}_\text{L}\text{L}_\text{L}}\mathbf{\overline{\Sigma}}^r_{\text{L}_\text{L}\text{D}_\text{L}\text{I}}  \mathbf{G}^{Tr}_{\text{II}} &\mathbf{\Sigma}^{{</>}}_{\text{I}\text{D}_{\text{R}}\text{I}} \mathbf{G}^{Ta}_{\text{II}} \mathbf{\overline{\Sigma}}^a_{\text{I}\text{D}_\text{L}\text{L}_\text{L}} \mathbf{g}^a_{\text{L}_\text{L}\text{L}_\text{L}},
\end{align}
\end{scriptsize}
where we defined the tunneling self-energy
\begin{align}
  \mathbf{\overline{\Sigma}}_{\text{L}_\text{L}\text{D}_\text{L}\text{I}} =  \mathbf{t}_{\text{L}_\text{L}\text{D}_\text{L}} \mathbf{\overline{g}}_{\text{D}_\text{L}\text{D}_\text{L}} \mathbf{t}_{\text{D}_\text{L}\text{I}}.
\end{align}
Inspection of~(\ref{eq:GTLlesser}) shows that this Green's function again accounts for equilibration in all components of the system, c.f.~(\ref{eq:GTkineticelaborate}). The first line in~(\ref{eq:GTLlesser}) contains the bare lesser/greater Green's function of the source lead, and thus accounts for direct equilibration in this lead. The second line describes equilibration after tunneling into the intermediate reservoir region. The third line describes tunneling from the source lead to equilibration in the drain lead, passing both quantum dots. Taking into account only one passage through each quantum dot as in section~\ref{sec:interactions}, we neglect the third term of~(\ref{eq:GTLlesser}). For the first term of~(\ref{eq:GTLlesser}) we employ
\begin{align}
&\left( \mathbf{I} + \mathbf{g}^r_{\text{L}_\text{L}\text{L}_\text{L}}\mathbf{\Sigma}^r_{\text{L}_\text{L}\text{D}_\text{L}\text{L}_\text{L}}\right) \mathbf{g}^{</>}_{\text{L}_\text{L}\text{L}_\text{L}}\left( \mathbf{I} + \mathbf{\Sigma}^a_{\text{L}_\text{L}\text{D}_\text{L}\text{L}_\text{L}} \mathbf{g}^a_{\text{L}_\text{L}\text{L}_\text{L}}\right) \nonumber \\ 
 &\qquad\qquad\qquad\qquad\qquad\qquad\to
\mathbf{g}^{</>}_{\text{L}_\text{L}\text{L}_\text{L}},
\end{align} 
compare~(\ref{eq:noreflections}).

\section{Energy integrals for finite range model interaction}
\label{app:EnergyIntegrals}

This appendix contains explicit expressions of integrals of~(\ref{eq:Xi}) for the finite range model interaction~(\ref{eq:GaussianModel}), required to evaluate all diagrams in which a charge carrier that has been excited from the Fermi sea passes through the detector.

Upon insertion of~(\ref{eq:GaussianModel}) into~(\ref{eq:Xi}), we find
\begin{align}
\label{eq:XiResult}
\Xi \left( \omega'' \right) &= \nu_0 \Bigg[ \frac{\Delta x}{\lambda^2 \omega''^2 + v^2} + \frac{2 i \lambda^2 v \omega''}{\left(\lambda^2 \omega''^2 + v^2\right)^2} \nonumber \\
 &\qquad \qquad \qquad- \frac{\lambda}{2} \frac{\exp \left( i \frac{\Delta x}{v} \left( i\frac{v}{\lambda} - \omega'' \right) \right)}{\left( i v -\lambda \omega'' \right)^2}   \Bigg],
\end{align}
in which we neglect the third term on the right in case $\Delta x \gg \lambda$.
Using this approximation, we have
\begin{align}
  &\Lambda\left( \omega'' \right) = \int d\omega'' \, \Xi \left( \omega'' \right)  \nonumber \\
 &= \nu_0 \left[\frac{\Delta x \arctan\left(\frac{\lambda  \omega'' }{v}\right)}{\lambda  v}-\frac{i v}{\lambda ^2 \omega''^2+v^2}\right],
\end{align}
\begin{align}
  &\Phi\left( \omega'' \right) = \int d\omega'' \, |\Xi \left( \omega'' \right)|^2 \nonumber \\
  &= \frac{1}{12} \nu_0^2 \Bigg[\frac{3 \left(\lambda ^2+2 \Delta x^2\right) \arctan\left(\frac{\lambda  \omega'' }{v}\right)}{\lambda  v^3}+\frac{2 \lambda ^2 \omega'' }{\left(\lambda ^2 \omega''^2+v^2\right)^2}\nonumber \\
  &\qquad \qquad-\frac{8 \lambda ^2 v^2 \omega''
   }{\left(\lambda ^2 \omega''^2+v^2\right)^3}+\frac{3 \omega''  \left(\lambda ^2+2 \Delta x^2\right)}{v^4+\lambda ^2 v^2 \omega''^2}\Bigg],
\end{align}
as well as 
\begin{scriptsize}
\begin{align}
\label{eq:XIXI2c}
&\Psi\left( \omega'' \right) = \int d\omega'' \, \Re \left[ \Xi^* \left( \omega'' \right) \Xi \left( \omega_L - \omega_R - \omega'' \right) \right] \nonumber \\
& = \frac{\nu_0^2}{\left(\lambda ^2 (\omega_L-\omega_R)^2+4 v^2\right)^3} \Bigg\{\arctan\left(\frac{\lambda  \omega'' }{v}\right) \big[\Delta x^2 \lambda ^4 (\omega_L-\omega_R)^4 \nonumber \\ 
&+8 v^4 \left(2 \Delta x^2-\lambda ^2\right)+2 \lambda ^2 v^2 \left(4
   \Delta x^2+3 \lambda ^2\right) (\omega_L-\omega_R)^2\big]\Big/\lambda  v \nonumber \\
&+\arctan\left(\frac{\lambda  (\omega'' -\omega_L+\omega_R)}{v}\right)\Big[\Delta x^2 \lambda ^4 (\omega_L-\omega_R)^4 \nonumber \\
&+8 v^4 \left(2 \Delta x^2-\lambda ^2\right)
+2 \lambda
   ^2 v^2 \left(4 \Delta x^2+3 \lambda ^2\right) (\omega_L-\omega_R)^2\Big] \Big/\lambda  v \nonumber \\ 
&  + \bigg\{2 v^2 (\omega_L-\omega_R) (-2 \omega'' +\omega_L-\omega_R) \left(\lambda ^2 (\omega_L-\omega_R)^2+4 v^2\right) \nonumber \\
  &  \times \Big[-\lambda ^4 (\omega_L-\omega_R)^2 \left(-3 \omega''^2+3 \omega''  (\omega_L-\omega_R)+(\omega_L-\omega_R)^2\right) \nonumber \\ 
   &+4 v^4+\lambda ^2 v^2 \Big(4 \omega''^2+4 \omega''  (\omega_R-\omega_L) \nonumber \\
   &+3 (\omega_L-\omega_R)^2\Big)\Big]\bigg/ \left(\lambda ^2 \omega''^2+v^2\right) \left(\lambda ^2 (\omega'' -\omega_L+\omega_R)^2+v^2\right) \nonumber \\
   &+\Big[\Delta x^2 \lambda ^4 (\omega_L-\omega_R)^6+32 v^6+8 v^4 \left(2 \Delta x^2+3 \lambda ^2\right)
   (\omega_L-\omega_R)^2 \nonumber \\
  & +4 \lambda ^2 v^2 \left(2 \Delta x^2+3 \lambda ^2\right) (\omega_L-\omega_R)^4\Big] \Big[\log \left(\lambda ^2 \omega''^2+v^2\right) \nonumber \\
  &-\log \left(\lambda ^2 (\omega'' -\omega_L+\omega_R)^2+v^2\right) \Big]\Big/\lambda ^2\bigg\}\bigg/(\omega_L-\omega_R)^3\Bigg\}. 
\end{align}
\end{scriptsize}

\bibliography{PRB-electron-spectrometer}

\end{document}